\RequirePackage{fix-cm}
\documentclass{svjour3}                     % onecolumn (standard format)
\smartqed  % flush right qed marks, e.g. at end of proof
\usepackage{graphicx}
\usepackage{multirow}
\usepackage[colorlinks=true, allcolors=blue]{hyperref}

\usepackage{cite}
\journalname{Experimental Astronomy}
\begin{document}

\title{Geant4 simulations of soft proton scattering in X-ray optics
%\thanks{Grants or other notes
%about the article that should go on the front page should be
%placed here. General acknowledgments should be placed at the end of the article.}
}
\subtitle{A tentative validation using laboratory measurements}

%\titlerunning{Short form of title}        % if too long for running head

\author{Valentina Fioretti         \and
             Teresa Mineo         \and
             Andrea Bulgarelli    \and
             Paolo Dondero    \and
             Vladimir Ivanchenko \and
             Fan Lei \and
             Simone Lotti          \and
             Claudio Macculi       \and
             Alfonso Mantero on behalf of the AREMBES collaboration
}

\authorrunning{V. Fioretti et al.} % if too long for running head

\institute{V. Fioretti \at
              INAF/IASF Bologna, Via P. Gobetti 101, 40129, Bologna, Italy \\
              \email{fioretti@iasfbo.inaf.it}           %  \\
%             \emph{Present address:} of F. Author  %  if needed
           \and                
   				T. Mineo \at
              INAF/IASF Palermo, Via Ugo La Malfa 153, 90146 Palermo, Italy
           \and
				A. Bulgarelli \at
              INAF/IASF Bologna, Via P. Gobetti 101, 40129, Bologna, Italy 
           \and
           P. Dondero \at
              SWHARD srl, Via Greto di Cornigliano, 6, 16152, Genova, Italy
%             \emph{Present address:} of F. Author  %  if needed
           \and       
             V. Ivanchenko \at
             CERN, Geneve 23, 1211, Switzerland \\ Geant4 Associate International Ltd, 9 Royd Terrace, Hebden Bridge HX7 7BT, United Kingdom
             \and
             F. Lei \at
             RadMod Research, 25 Parkway, Camberley, Surrey, GU15 2PD, United Kingdom
 				\and
   				S. Lotti \at
              INAF/IAPS, Via del Fosso del Cavaliere, 100, 00133 Roma, Italy
%             \emph{Present address:} of F. Author  %  if needed
           \and        
   				C. Macculi \at
              INAF/IAPS, Via del Fosso del Cavaliere, 100, 00133 Roma, Italy
%             \emph{Present address:} of F. Author  %  if needed
           \and        
   				A. Mantero \at
              SWHARD srl, Via Greto di Cornigliano, 6, 16152, Genova, Italy
%             \emph{Present address:} of F. Author  %  if needed
           }

\date{Received: date / Accepted: date}
% The correct dates will be entered by the editor

\maketitle

\begin{abstract}
Low energy protons ($< 300$ keV) can enter the field of view of X-ray space telescopes, scatter at small incident angles, and deposit energy on the detector. This phenomenon can cause intense background flares at the focal plane decreasing the mission observing time (e.g. the XMM-Newton mission) or in the most extreme cases, damaging the X-ray detector. A correct modelization of the physics process responsible for the grazing angle scattering processes is mandatory to evaluate the impact of such events on the performance (e.g. observation time, sensitivity) of future X-ray telescopes as the ESA ATHENA mission.
The Remizovich model describes particles reflected by solids at glancing angles in terms of the Boltzmann transport equation using the diffuse approximation and the model of continuous slowing down in energy.
For the first time this solution, in the approximation of no energy losses, is implemented, verified, and qualitatively validated on top of the Geant4 release 10.2, with the possibility to add a constant energy loss to each interaction. This implementation is verified by comparing the simulated proton distribution to both the theoretical probability distribution and with independent ray-tracing simulations. Both the new scattering physics and the Coulomb scattering already built in the official Geant4 distribution are used to reproduce the latest experimental results on grazing angle proton scattering. At 250 keV multiple scattering delivers large proton angles and it is not consistent with the observation. Among the tested models, the single scattering seems to better reproduce the scattering efficiency at the three energies but energy loss obtained at small scattering angles is significantly lower than the experimental values.
%Single scattering and Remizovich efficiency in reflecting protons is consistent with the overall experimental behaviour, but for small angles the simulation gives higher efficiencies
%The overall angular distribution is well reproduced by all the models at 500 and 1000 keV, except for very low scattering angles, where the Geant4 simulation gives higher scattering efficiency. At 250 keV, the multiple scattering translates in large scattering angles and it is not consistent with the observation. 
In general, the energy losses obtained in the experiment are higher than what obtained by the simulation.
%, with the multiple scattering better reproducing the data set at 500 and 1000 keV. 
The experimental data are not completely representative of the soft proton scattering experienced by current X-ray telescopes because of the lack of measurements at low energies ($<200$ keV) and small reflection angles, so we are not able to address any of the tested models as the one that can certainly reproduce the scattering behavior of low energy protons expected for the ATHENA mission. We can, however, discard multiple scattering as the model able to reproduce soft proton funnelling, and affirm that Coulomb single scattering can represent, until further measurements at lower energies, the best approximation of the proton scattered angular distribution at the exit of X-ray optics.

%, and consequently driving the design of a magnetic diverter. 
%Among the tested models, the single scattering seems to better reproduce the scattering efficiency at the three energies but significant lower energy losses are obtained at small scattering angles.

% \PACS{PACS code1 \and PACS code2 \and more}
% \subclass{MSC code1 \and MSC code2 \and more}
\end{abstract}

\section{Introduction}
Charged particles can pose a significant radiation threat to the on-board electronic systems of X-ray space missions, depending on the telescope orbit. This phenomenon is particularly complex for grazing incident X-ray telescopes as the NASA Chandra X-ray Observatory (CXO)\cite{2000SPIE.4012....2W} and the ESA XMM-Newton \cite{2001A&A...365L...1J} telescopes, launched in July and September 1999 respectively, and currently operating in a highly eccentric orbit that crosses the radiation belt in the $(5-10)\times10^{3}$ km altitude range. Both telescopes carry Wolter-I type mirrors to focus X photons through grazing angle reflection to the detection plane. The capability of X-ray optics to focus electrons was already known before the launch of the two missions. For this reason, X-ray telescopes were equipped with magnetic diverters that deflect the electron paths outside the detection plane (see e.g \cite{wil00}). However, the loss of charge transfer efficiency suffered by the Chandra Advanced CCD Imaging Spectrometer (ACIS \cite{2003SPIE.4851...28G}) Front-Illuminated (FI) CCDs after the first radiation belt passages revealed that low energy protons can also be reflected by the X-ray mirror shells and reach the focal plane. With energies between 1-300 keV, these so-called ``soft protons" caused serious damages to the FI CCD at the focal plane of CXO and the overall mission performance of XMM-Newton. In fact, XMM-Newton's filter wheel completely blocks the EPIC \cite{1996SPIE.2808..402V} field of view when crossing the radiation belts protecting the detectors from damage. Unfortunately above the radiation belt limit, where the instruments are fully operative, the soft proton funnelling is still observed by the XMM-Newton detectors in the form of sudden increases in the background level. Such soft proton flaring events can prevail over the quiescent background level up to orders of magnitude, affecting 30-40\% of XMM-Newton observing time \cite{2007A&A...464.1155C}. Soft proton flares are extremely unpredictable in duration, lasting from $\sim100$ s to hours \cite{del04}, and generate an average count rate, in all three CCDs, of 2-2.5 prot. cm$^{-2}$ s$^{-1}$ \cite{lum02}. Several studies (see e.g. \citenum{2007A&A...464.1155C}) have proven the solar origin of this damaging background. Missions operating in low Earth orbit (e.g. Swift \cite{2005SSRv..120..165B}, Suzaku \cite{2007PASJ...59S...1M}) do not suffer from soft proton flares, thanks to the geomagnetic shield.
Soft protons populating the outer magnetosphere, including the magnetotail, the magnetosheath and the solar wind, both in the form of a steady flux and violent Coronal Mass Ejections, can instead increase the X-ray residual background level and even threaten the observation itself \cite{bri00}.
\\
Future X-ray focusing telescopes operating outside the radiation belts (e.g. the ESA Athena\cite{2013arXiv1306.2307N} mission and the eROSITA\cite{2014SPIE.9144E..1TP} instrument on-board the Russian Spektr-RG observatory, both to be placed in L2 orbit) will also be affected by soft proton contaminations. 
%Monte Carlo simulations are mandatory to evaluate the impact of such events and accordingly design shielding solutions (e.g., a magnetic diverter) without limiting the sensitivity of the instruments. 
\\
A validated physics model to describe the angular and energy proton distribution at the exit of the optics is mandatory for a correct evaluation of the impact of soft proton events to the mission performance.
Despite many solutions proposed so far to explain the physics interaction behind the soft proton grazing angle scattering (see e.g. \cite{spi08}), the difficulty in implementing dedicated models prevented Monte Carlo codes from developing physically sound models.  
We implement in the Geant4 release 10.2 both the Firsov distribution following the work of \cite{lei04} and for the first time the bivariate distribution described by Remizovich in its elastic approximation (see Sec. \ref{sec:impl}).
%, to take into account both the polar and the azimuthal distribution of protons after low angle interactions. 
After a dedicated verification of the implementation performed comparing the results with the analytical model and the ray-tracing simulation, we use the latest scattering measurements obtained at 250, 500, and 1000 keV proton energy by \cite{2015ExA....39..343D} on eRosita shell samples to accurately compare the measured energy, angular, and intensity distribution of protons with the ones predicted from both the new models and the Geant4 default library.
The physics validation of a soft proton scattering model is achieved if we are able to reproduce, within an acceptable uncertainty level, the experimental data. 
%, and require the accurate characterization of (i) the distribution of the soft proton population, (ii) the mirror-proton physics interaction at play, and (iii) the effect on the focal plane. Despite many solutions proposed so far to explain the physics interaction behind the soft proton grazing angle scattering, the difficulty in obtaining accurate laboratory measurements\cite{2015ExA....39..343D} prevents ray-tracing codes from implementing physically-reliable models. 
%We implement in the Geant4 release 10.2 both the Firsov distribution following the work of \cite{lei04} and for the first time the bivariate distribution described by Remizovich in its elastic approximation, to take into account both the polar and the azimuthal distribution of protons after low angle interactions. After a dedicated verification of the implementation by comparing the results with the analytical model and the ray-tracing simulation, we simulate the experimental set-up of \cite{2015ExA....39..343D} to accurately compare the expected energy, angular, and intensity distribution of protons with the ones predicted from the Remizovich elastic model and the current Coulomb scattering processes provided by the Geant4 library.
%The physics validation of a soft proton scattering model is achieved if we are able to reproduce, within an acceptable uncertainty level, the experimental data. 

\section{Experimental data}
The experiment of \cite{2015ExA....39..343D} evaluates the scattering efficiency, in sr$^{-1}$, and the energy loss of protons at 250, 500, and 1000 keV interacting with eRosita shell samples at glancing angles in the $0.3^{\circ} - 1.2^{\circ}$ range. 
The general set-up consists of a proton beam line produced by a ion accelerator facility hitting the X-ray mirror shell sample at different angles and then collected by a shiftable proton detector:
\begin{itemize}
\item the incident angle has a precision, in terms of tilt angle, of $0.006^{\circ}$;
\item the eRosita target shell is composed by a Nickel substrate of 270 $\mu m$ coated by 50 nm of Gold and different sample sizes, ranging from a length of 10 cm to 12 cm, were used with no impact on the measurements;
\item the proton detector consists of a 8 mm wide Silicon surf barrier detector characterized by a detection efficiency of almost 100\% and an overall accuracy within $\pm10$ keV.
\end{itemize}
%\subsection{Scattering efficiency}\label{sec:scatt}
The 1.2 mm aperture at the proton detection point, placed at a distance of 933 mm from the target, defines a proton collection solid angle $\Omega\simeq1.3$ $\mu$sr. The scattering angle is defined in the reference as $\theta + \theta_{0}$, the sum of the scattering and incident polar angles. The proton scattering efficiency $\eta(\phi, \theta)$, in sr$^{-1}$,  is computed by dividing the number of detected protons N$_{\rm det}$ by the number of incident protons in the target, N$_{\rm inc}$, and the aperture solid angle $\Omega$:
\begin{equation}\label{eq:eff}
\eta(\phi, \theta) = \frac{\rm N_{\rm det}}{N_{\rm inc}\times\Omega}\: .
\end{equation}
The incident angles, the resulting scattering efficiency values and the energy losses are taken as reference values in the validation test of Sec. \ref{sec:val}.
\section{Investigated theoretical models}

\subsection{Remizovich model}
The analytical model of \cite{1980JETP...52..225R, 1983RadEffect} describes particles reflected by solids at glancing angles in terms of the Boltzmann transport equation using the diffuse approximation and the model of continuous slowing down in energy. According to their model, the proton energy loss is peaked at about 10-20\% of the initial energy E$_{0}$, i.e. ranges from 5-10 keV at 50 keV to 50-100 keV at 500 keV.
The mathematical form of the Remizovich model in its elastic approximation is much simpler than the full model and does not depend on the physical properties on the reflecting material. Besides that, the energy loss can be treated, in first approximation, as constant, and for this reason we use the elastic Remizovich approximation to model the angular distribution of protons after reflection at glancing angles.
%, and subtract “a posteriori” a Gaussian distributed energy amount to the reflected protons. 
In the present simulations we added to the Remizovich scattering a $3\pm0.7$ keV energy loss, as in \cite{lei04}. This is the most probable energy loss found in past laboratory measurements of grazing proton scattering on different surfaces in the 30 - 710 keV energy range \cite{1987PhRvB..36....7K}. 
\\
The elastic approximation of the Remizovich solution takes the form:
\begin{eqnarray}\label{eq:rem}
\rm W_{\rm el}(\Psi, \chi) = \frac{1}{12\pi^{2}\Psi^{1/2}}\left[ \frac{\omega^{4}}{1 + \omega^{2}} + \omega^{3}arctan(\omega)\right], \\
\omega = \left\lbrace \frac{3\Psi}{\Psi^{2}-\Psi+1+(\chi/2)^{2}}\right\rbrace^{1/2}.
\end{eqnarray}

\subsection{Firsov model}
The elastic probability distribution of Eq. \ref{eq:rem} integrated all over the azimuthal angles becomes the formulation obtained by Firsov to describe the reflection of fast ions from a dense medium at glancing angles:
\begin{equation}\label{eq:firsov}
\rm W(\Psi) = \frac{3}{2\pi}\frac{\Psi^{3/2}}{1 + \Psi^{3}}.
\end{equation}
The Firsov formula was used in \cite{lei04} in the assumption that all protons are scattered at $\phi=0$, i.e. continue their path along the x-axis of Fig. \ref{fig:slab} (right panel). In order to compare our results with past simulations we also implemented Eq. \ref{eq:firsov} with $\phi = \phi_{0} = 0$. We will generally refer to this implementation as Firsov.

\subsection{Single and multiple scattering model}
In the single or multiple Coulomb scattering, when a charged particle traverses a medium, it undergoes one or more elastic scatterings due to Coulomb interactions with the electron field of the nuclei, as described by the Rutherford cross section. 
%In case of many small angle scatters, the multiple Coulomb angular distribution is usually described by the theory of Moli\`ere \cite{moliere}, giving a width $\delta_{0}$ of the angular distribution, projected on the scattering plane, of:
%\begin{equation}\label{eq:mol}
%\delta_{0} = \frac{13.6 \: \rm MeV}{\beta \rm c p}\sqrt{\frac{\rm t}{\rm X_{0}}}\left(1 + 0.038 \:\rm ln \frac{\rm t}{\rm X_{0}}\right)\:,
%\end{equation}
%where p, $\beta$c, z, and $\rm t/X_{0}$ are the momentum, velocity, charge of the incident particle, and the thickness of the s%cattering medium in radiation lengths respectively.
\\
For grazing incident angles, protons can interact with the the nuclei at the material edge and escape after one or more interactions, with the effect of being scattered by the target, often with an enhanced deflection angle with respect to the incident one if multiple scattering is involved.
\\
Multiple scattering was the first model to be proposed as the one responsible for soft proton funneling by X-ray optics.

\section{Geant4 simulation set-up}\label{sec:setup}
\begin{figure}
\center
  \includegraphics[trim={0.5cm 0 0 0},clip, width=0.49\textwidth]{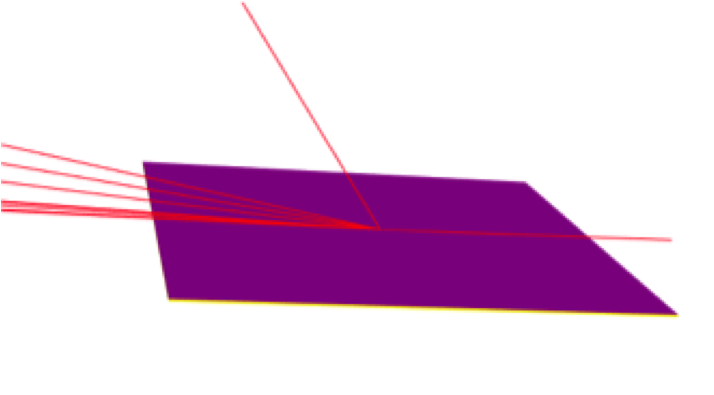}
  \includegraphics[width=0.49\textwidth]{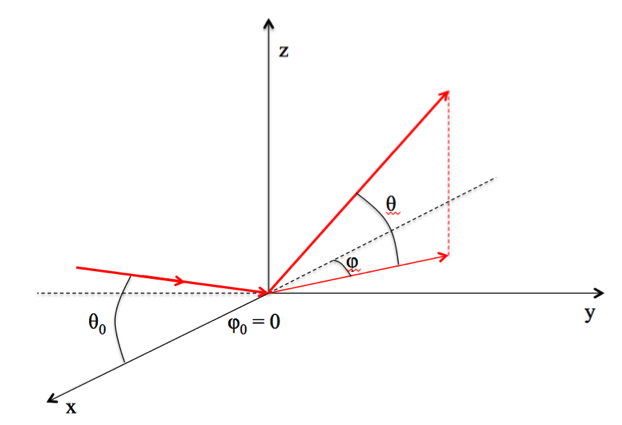}
\caption{\textit{Left panel:} The eRosita Au-coated Ni shell portion is approximated by two planar slabs. The red lines show the proton trajectory before (right side) and after (left side) the interaction. \textit{Right panel:} The polar $\theta$ and azimuthal $\phi$ angle definition used for the Remizovich and Firsov formulation, with the reflecting surface placed in the x-y plane and the proton trajectory highlighted in red.}
\label{fig:slab}     
\end{figure}

The Geant4 \cite{g4_1, g4_2, 2016NIMPA.835..186A}  Monte Carlo toolkit is a C++ based particle transport code, initially developed by CERN for the simulation of high energy experiments at particle accelerators and then extended to lower energy ranges, i.e. the X and Gamma-ray domain. Geant4 has become the standard tool used by many space agencies (e.g. ESA) in the simulation of the background and instrument performance of all major X-ray space telescopes (e.g., Chandra, XMM-Newton, Suzaku, Athena, eROSITA).
\\
BoGEMMS (Bologna Geant4 Multi-Mission simulator) is a Geant4-based simulation tool developed at the INAF/IASF Bologna \cite{2012SPIE.8453E..35B, 2014SPIE.9144E..3NF} for the evaluation of the scientific performance (e.g. background spectra, effective area) of X-ray and Gamma-ray space missions. It allows to interactively set the geometrical and physical parameters recording the interactions in \textit{FITS} and \textit{Root} format output files and filtering the output as a real observation in space, to finally produce the background detected count rate and spectra. 
The BoGEMMS framework and the Geant4 release 10.2 are used throughout the activity presented in this paper, with a selection cut for all volumes of 1 nm. Unless otherwise specified, we use Geant4 reference physics list \textit{QGSP\_BERT\_HP} with the \textit{opt3} electromagnetic physics list on top.
%Unless otherwise specified, the default electromagnetic physics uses the \textit{G4EmStandardPhysics\_option3} list and the default hadron physics uses the \textit{QGSP\_BIC\_HP} list. 
Given the involved energies, the simulation does not use the latter list, but it is present to check that the Geant4 simulation correctly selects the new proton physics.
Since the validation of proton scattering in Geant4 is achieved by comparing the simulation with the experiment of \cite{2015ExA....39..343D}, the mass model of the reflecting surface used for both the verification and validation tests (see Figure \ref{fig:slab}, left panel) approximates the eRosita X-ray shell sample used in the experiment: a planar slab of Nickel, 270 $\mu$m thick, coated by 50 nm of Gold.
\\
A new physics list \textit{G4SoftProtonPhysics} is added for the handling of the two new models describing the scattering of protons at grazing angles: the Remizovich elastic approximation and the Firsov azimuthal elastic integration. 
Thanks to the BoGEMMS configuration framework, the user can set at run-time the proton energy and incident angular range where to apply the new models. 
Models can also be combined and used in the same simulation using different energy or incident angle ranges. The probability for the proton to undergo a Firsov or Remizovich interaction is arbitrary set to 100\%, if the proton matches the energy and angle range of applicability. This assumption is due to the goal of the present study, i.e. low energy and low angle proton scattering.

\section{Physics implementation and verification}\label{sec:impl}
Among the many physical processes that have been proposed to describe the scattering of soft protons by X-ray optics, only the Coulomb single scattering model is currently available in the official Geant4 toolkit \cite{nar01}. Geant4 simulations of the soft proton scattering by the XMM-Newton X-ray optics were updated by implementing the Firsov angular scattering distribution on top of the Geant4 version 9.1, with the addition of a constant small energy loss \cite{lei04}. The two new Geant4 physics classes developed for this purpose, G4FirsovSurface and G4FirsovScattering, were not included in the official release of the Geant4 toolkit.
%Recent proton scattering measurements using eRosita mirror shells \cite{2015ExA....39..343D} have shown that, although the Firsov formulation reproduces the overall behavior, systematic shifts in the efficiency and energy losses are present. The Firsov formula is however the azimuthal integration of the proton distribution described by Remizovich and collaborators in \cite{1980JETP...52..225R}. 
%Using the BoGEMMS framework, the user can set at run-time both the mean energy loss and its standard deviation.
\\
We define as $\theta_{0}$ and $\phi_{0}$ the incident polar and azimuthal angles with respect to the reflecting surface and as $\theta$ and $\phi$ the polar and azimuthal angles after the proton scattering. For each proton reaching the surface at glancing angle, the x-axis of the Cartesian reference system is placed along the proton trajectory, so that  $\phi_{0}$ is always zero. Figure 2 shows the angular system used to describe the reflection, with the reflecting material placed in the x-y plane and the proton trajectory highlighted in red. 
Following the formalism of \cite{1980JETP...52..225R}, we introduce the dimensionless variables $\Psi = \theta/\theta_{0}$, $\chi= \phi/\theta_{0}$ to express the proton angular distribution. 
\\
Given the simple slab geometry, in both Remizovich and Firsov model implementations no check is inserted on the reflecting surface material, the only requirement for the soft proton scattering to be activated is to match the incident energy and angles. The dependence on the material properties (e.g. density, atomic number) will be inserted in future releases of the physics classes.

\subsection{Firsov}\label{sec:firsov}

The range of possible scattering polar angles, from $0^{\circ}$ to $90^{\circ}$, is divided in a limited number of discrete possible values, and the resulting proton angle after scattering is randomly picked up among the list of discrete bins following the analytical probability distribution. The number of bins used to divide the scattering angle range has a direct impact on the resolution of the scattering angles. Since our final goal is to reproduce the experimental results of \cite{2015ExA....39..343D}, the uncertainty level in the scattering angle distribution must be lower than the angular dimension of the proton detection area used in the laboratory measurements. In this case, a 1.2 mm side aperture at a focal distance of 933 mm from the center of the X-ray shell translates into an angular resolution of $\sim0.07^{\circ}$, meaning at least 1000-2000 discrete bins of the polar angle range. For the present simulation, we use a value of 10000.
\begin{figure}
\center
  \includegraphics[width=0.49\textwidth]{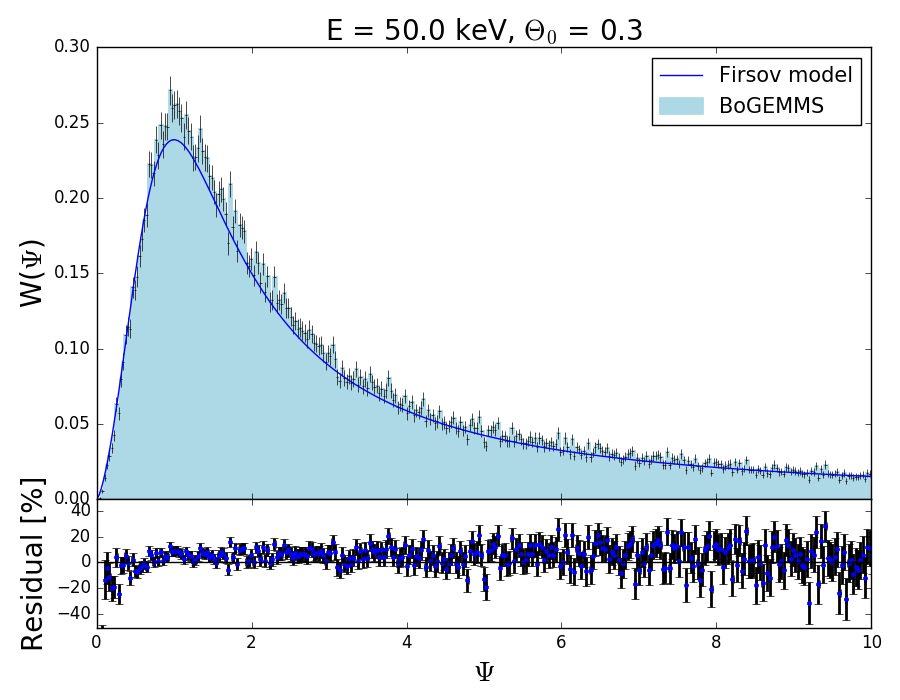}
  \includegraphics[width=0.49\textwidth]{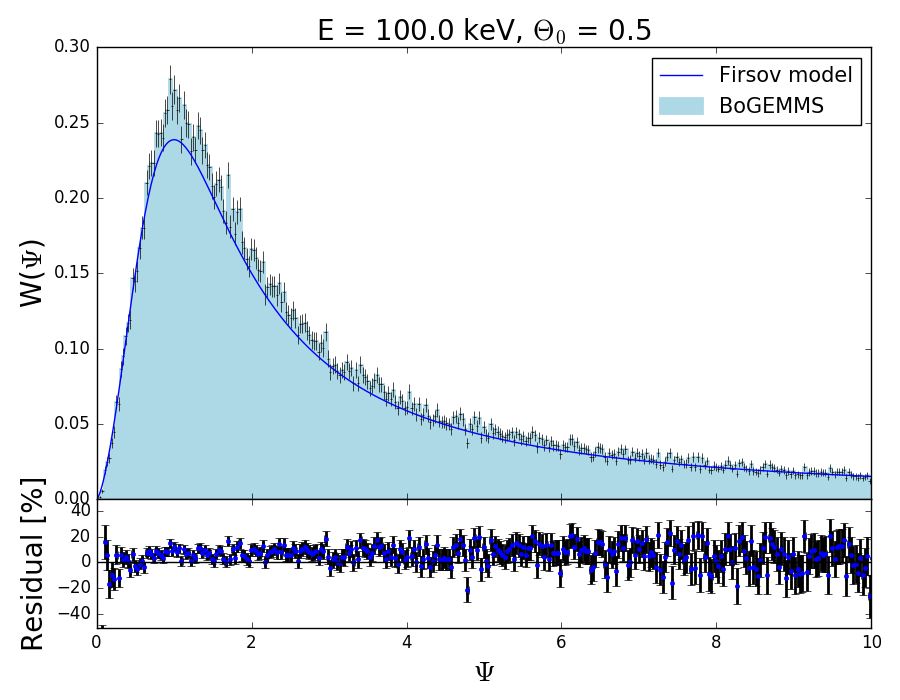}
\caption{Comparison between the Firsov angular distribution W($\Psi$) of scattered protons obtained by the BoGEMMS simulation (light blue area) and the analytical model (dark blue line) for an incident energy and polar angle of (50 keV and $0.3^{\circ}$, left panel) and (100 keV and $0.5^{\circ}$, right panel).}
\label{fig:firsov_ver}     
\end{figure}
\\
The simulation set-up for the verification phase consists of a proton point source in correspondence of the selected incident polar angle $\theta_0$. Although no dependence is expected on the initial proton energy, we use two different initial energies (50 and 100 keV) to check the correct setting, at run-time, of the proton scattering range of applicability. The resulting probability distribution W as a function of the dimensionless polar variable $\Psi $ is shown in Fig. \ref{fig:firsov_ver}, for an incident angle of $0.3^{\circ}$ (left panel) and $0.5^{\circ}$ (right panel). We are able to reproduce the Firsov distribution within $\sim10\%$ of the analytical values. 

\subsection{Remizovich}
\begin{figure}[h!]
\center
  \includegraphics[width=0.99\textwidth]{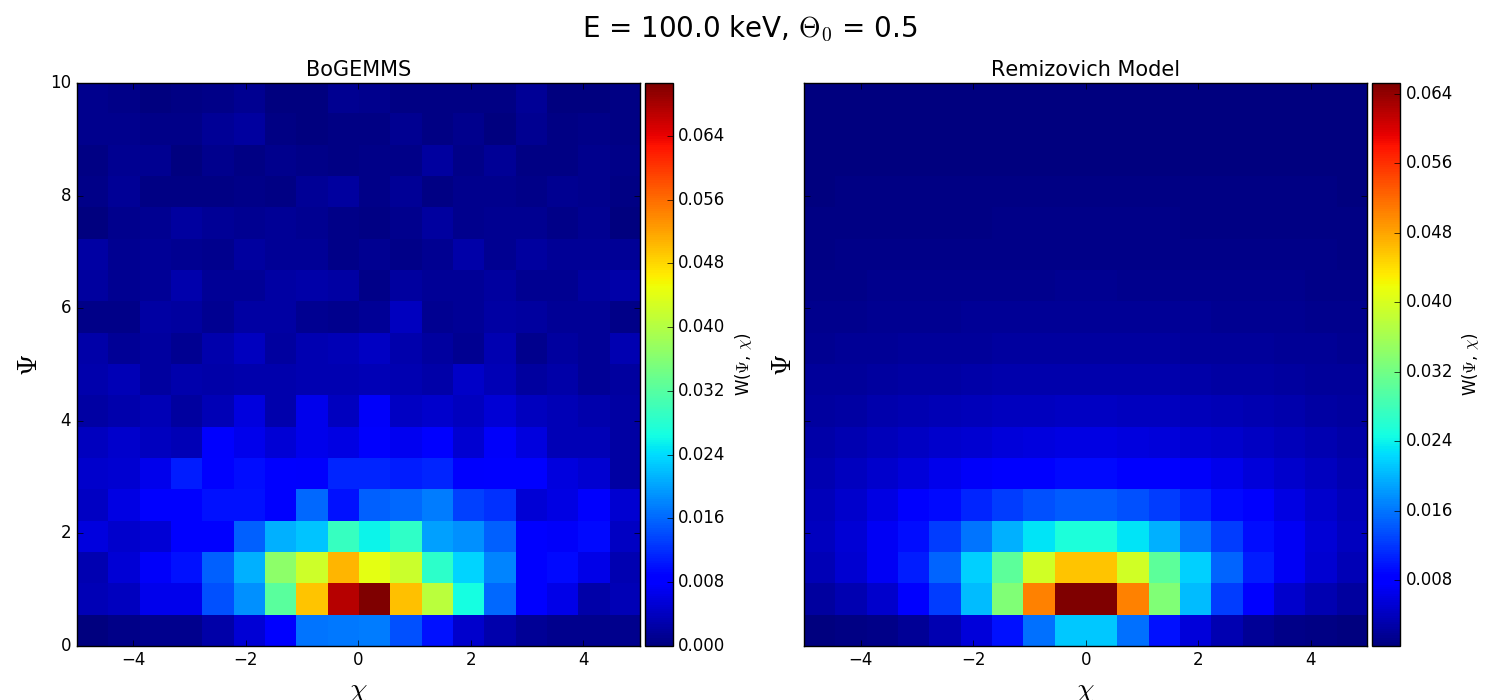}
\caption{Comparison between the bi-dimensional Remizovich angular distribution W($\Psi$, $\chi$) of scattered protons obtained by the BoGEMMS simulation (left panel) and the analytical model (right panel) for an incident energy and polar angle of 100 keV and $0.5^{\circ}$.}
\label{fig:remiz_map}     
\end{figure}

The Remizovich model describes the distribution in both $\theta$ and $\phi$, or $\Psi$ and $\chi$, of the scattered protons. The Geant4 implementation requires dividing in discrete bins both the polar and azimuthal ranges of the scattering angle, from $0^{\circ}$ to $90^{\circ}$ and in the $\pm90^{\circ}$ range respectively. As described in the previous section for the Firsov case, the number of discrete bins, 3000 for the present simulations, is decided according to the required angular distribution. The binned probability distribution, in the $\Psi-\chi$ parameter space, obtained from the BoGEMMS simulation and the analytical model is plotted in Fig. \ref{fig:remiz_map}, with the color bar showing the value of the binned distribution W($\Psi$, $\chi$).
\begin{figure}
\center
  \includegraphics[width=0.49\textwidth]{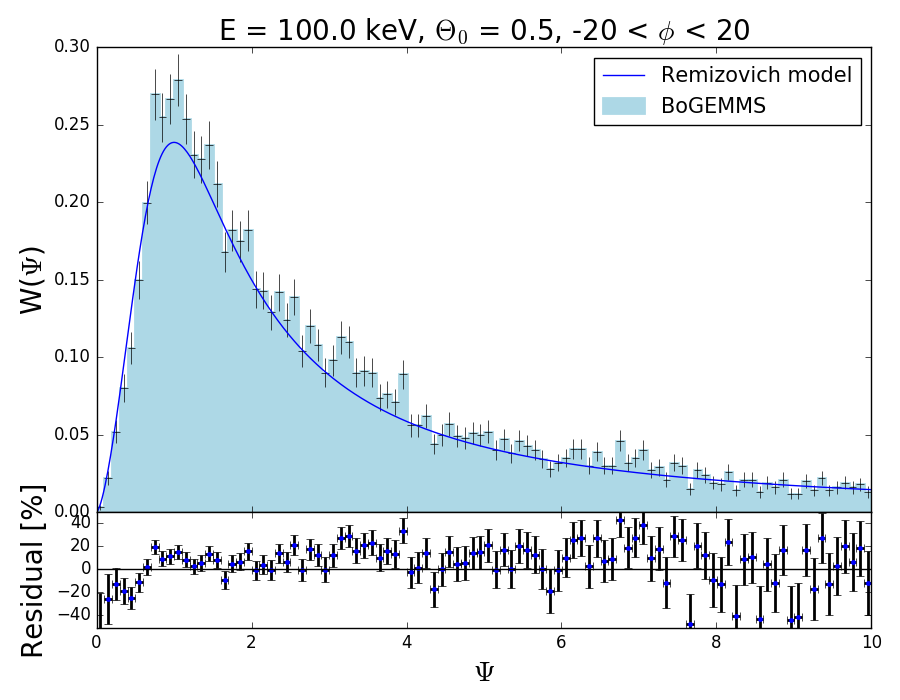}
  \includegraphics[width=0.49\textwidth]{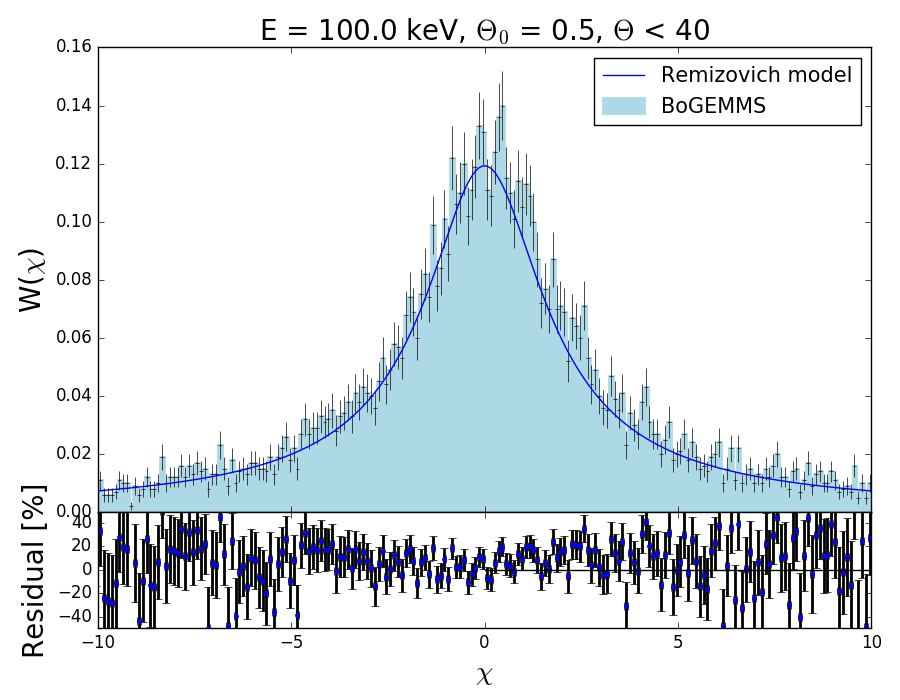}
\caption{Linear comparison of the integrated Remizovich distribution, for a polar incident angle of $0.5^\circ$, by integrating the azimuthal angle range $\psi=\pm20^{\circ}$ (left panel) and the polar angle $0^{\circ} < \theta < 40^{\circ}$ (right panel).}
\label{fig:remiz_ver}     
\end{figure}
\begin{figure}
\center
  \includegraphics[width=0.6\textwidth]{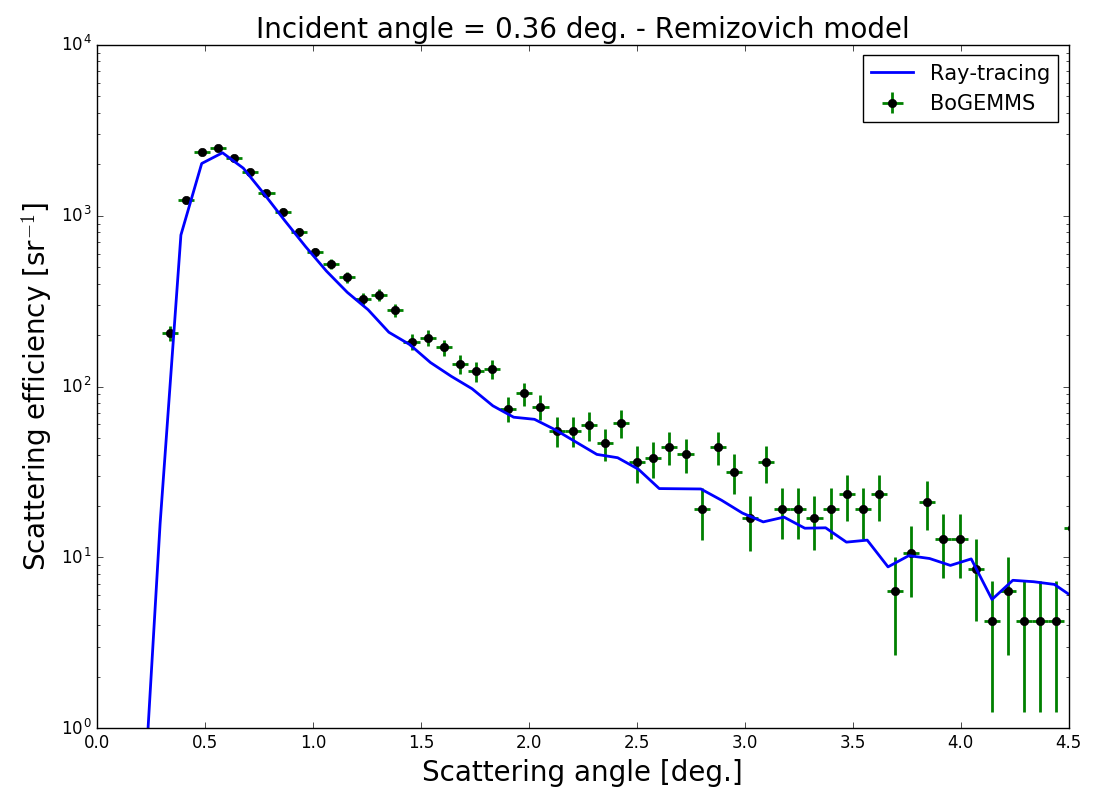}
\caption{Comparison of Remizovich induced scattering efficiency obtained with the two independent Geant4 (crosses) and ray-tracing (continuous line) simulations.}
\label{fig:g4_ray}     
\end{figure}
Fig. \ref{fig:remiz_ver} shows the linear comparison between the BoGEMMS and the Remizovich model obtained by integrating over a range of scattering azimuthal angles  $\psi=\pm20^{\circ}$ (left panel) and polar angles $0^{\circ} < \theta < 40^{\circ}$ (right panel). The model is reproduced within a maximum uncertainty of $\sim20\%$.
The error bars are $1\sigma$ Poisson fluctuations of the number of detected events for each bin and depend on the total number of emitted protons. 
\\
The Remizovich implementation in the Geant4 code has been compared to the ray-tracing simulation of \cite{INAF-XIFU-TM-2015-1} for an independent verification. Contrary to the Geant4 particle transport code, a ray-tracing Monte Carlo simulator follows each particle from the mirror to the focal plane by applying the reflection model of the optics design. The present ray-tracing code, used in the design of the MECS on board of BebboSAX \cite{1994SPIE.2279..101C} and for the calibration of the Swift XRT effective area \cite{2006AIPC..836..664C}, has been modified to account for the reflection of soft protons by the ATHENA pore optics. As shown in Fig. \ref{fig:g4_ray}, the scattering efficiency (see Sec. \ref{sec:scatt} for a detailed description) predicted by the two simulators using the Remizovich solution is in very good agreement, confirming the proper construction of the process in Geant4.

\subsection{Single and multiple scattering}

In Geant4 10.2 there are two models describing the scattering of protons \cite{2010JPhCS.219c2045I}, applied here to grazing angle interactions: the Urban model of multiple scattering (\textit{G4UrbanMscModel}) used in the electromagnetic physics list \textit{opt3} and a combination of multiple scattering {\em G4WentzelVIMscModel} and single scattering model {\em G4eCoulombScatteringModel} used by default and for the \textit{opt4} physics list. These combinations of models are working simultaneously: one of the two models is called depending on the step and the relative probability. Small scattering angles are sampled by the multiple scattering model, large scattering angles by single scattering. These model are coherent, because they are using the same elastic cross section of Wentzel \cite{wentzel}. In the {\em G4EmStandardPhysicsSS} physics list instead only the single scattering model is defined, for any scattering angle.
%In Geant4 10.2 there are two models describing the scattering of protons \cite{2010JPhCS.219c2045I}, applied here to grazing angle interactions: the Urban model of multiple scattering (\textit{G4UrbanMscModel}) used in the electromagnetic physics list \textit{opt3} and a combination of multiple and single scattering used in the \textit{G4EmStandardPhysicsSS} physics list based on the WentzelVI model (\textit{G4WentzelVIModel}). 
\\
The Urban model uses empirical parameterisations to sample large scattering angles, which may not be accurate for all cases. It requires validation for each particular setup and often extra step limitations to get better agreement with the data. The WentzelVI multiple scattering model is more accurate but also requires checks of the optimal step limit for concrete uses cases. The single scattering model may be used out of the box and does not need a special tuning of step limits.
\\
These three types of physics lists simulate different trajectories of grazing protons in the absorber, and for this reason energy losses computed by Geant4 ionisation models can be different.
%Urban uses model functions to determine both the angular and spatial distribution after a simulation step. Being a condensed model, the Urban algorithm simulates the final particle properties, net energy loss, displacement, and change of direction of the charged particle after multiple scattering repetitions. This approach reduces in general the accuracy of the process but allows the efficient simulation of large number of collisions.
%The mixed simulation algorithm is based on the Wentzel screened Rutherford cross section of Wentzel \cite{wentzel}, and is the current default for the electromagnetic physics list \textit{opt4}. The choice of single or multiple scattering is dynamic, it depends on particle momuntum, step size, and the distance to the geometry boundary of a step.
%The Coulomb multiple scattering (MSC) is applied in its standard configuration as provided by the \textit{G4EmStandardPhysics\_option3} (Geant4 10.2) and it uses the Urban model (\textit{G4UrbanMscModel}). Contrary to the Moli\`ere theory that only provides the angular distribution, Urban uses model functions to determine both the angular and spatial distribution after a step. Being a condensed model, the Urban algorithm simulates the final particle properties, net energy loss, displacement, and change of direction of the charged particle at the end of the track. This approach reduces in general the accuracy of the process but allows the efficient simulation of large number of collisions.

\section{Comparison with real data}\label{sec:val}
\begin{figure}
\center
  \includegraphics[trim={0cm 0.1cm 0.1cm 0},clip, width=0.6\textwidth]{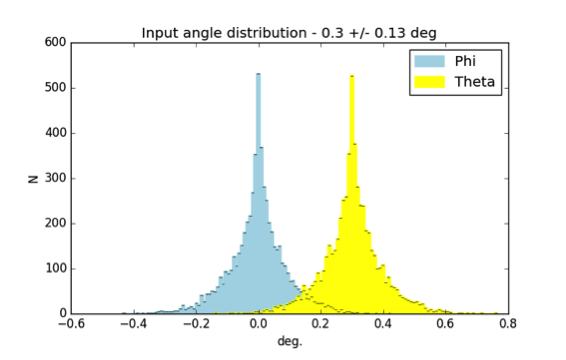}
\caption{The polar, in yellow, and azimuthal, in light blue, angular distribution of the proton beam profile at an incident angle of $0.3^{\circ}$ and a standard deviation of $0.13^\circ$.}
\label{fig:beam}     
\end{figure}

The uncertainty in the incident proton angle is simulated using as proton source a beam profile with a standard deviation equal to the angle error, as shown in Fig.~\ref{fig:beam}.
In addition to the Firsov and Remizovich formula, the proton scattering is simulated using the following available physics processes:
\begin{itemize}
\item Multiple Coulomb scattering, provided by the default \textit{G4EmStandardPhysics\_option3} electromagnetic physics list (\textit{opt3});
\item Multiple Coulomb scattering, provided by the \textit{G4EmStandardPhysics\_option4} electromagnetic physics list (\textit{opt4});
\item Single Coulomb scattering, provided by the \textit{G4EmStandardPhysicsSS} electromagnetic physics list.
\end{itemize}
All processes have been used with the standard settings provided by the Geant4 10.2 release.

\subsection{Scattering efficiency}\label{sec:scatt}

Since the scattering angular distribution is not isotropic, it is important in the simulation to collect the protons in the same solid angle used in the experiment. The results presented here are obtained by selecting only the scattered protons within the azimuthal range $\phi=\pm0.037^{\circ}$, the detection aperture subtended angle, and dividing the polar range in discrete bins in order to obtain, for each bin, the aperture solid angle $\Omega$. The number of protons in each bin is N$_{\rm det}$ of Eq. \ref{eq:eff}. Being both Firsov and Remizovich models elastic - the resulting scattering efficiency does not depend on the initial proton energy - we compare the output of the simulation with respect to experimental results for the three proton energies.
\begin{figure}
\center
  \includegraphics[width=0.8\textwidth]{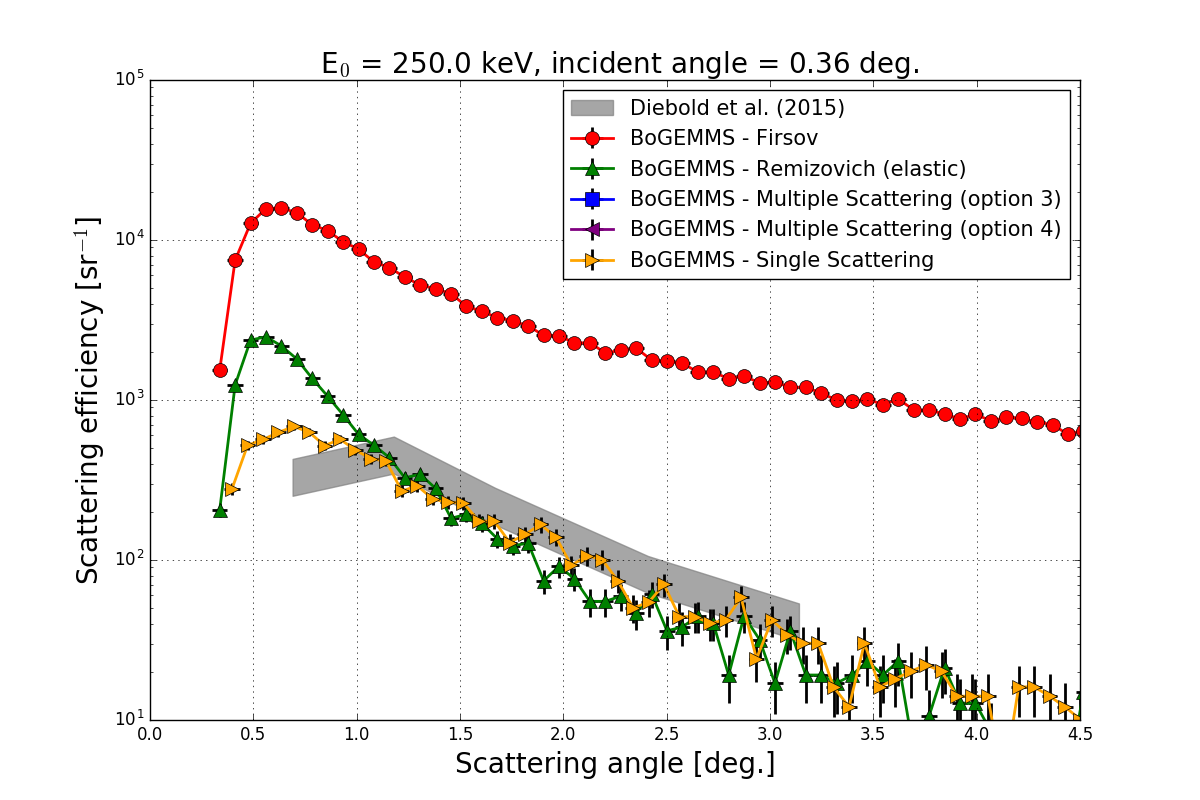}
\caption{Scattering efficiency at E$_{0}$ = 250 keV for an incident angle of $0.36^{\circ}$. The grey area shows the experimental results. Multiple scattering, both in \textit{opt3} and \textit{opt4}, results in higher scattering angles.}
\label{fig:eff_250_1}     
\end{figure}
The scattering efficiency for an incident proton energy of 250 keV and incident angle of $0.36^{\circ}$ is shown in Fig. \ref{fig:eff_250_1}.  The Firsov formula implemented with $\phi=0$ has the effect of focusing all protons in the detection area with the results of overestimating by more than an order of magnitude the scattering efficiency. Multiple scattering, for both \textit{opt3} and \textit{opt4}, causes at 250 keV scattering angles larger than the values obtained in the experiment: as shown in Fig. \ref{fig:MSC} (left panel), the scattering polar angle peaks in the $10^{\circ}-20^{\circ}$ range. 
\\
If small incident angles are considered, the Remizovich and the single scattering (SS) both well reproduce the proton angular distribution at scattering angles higher than $1^{\circ}$. Near the specular reflection ($\sim0.7^{\circ}$) both models give higher values, with the SS inducing a scattering efficiency closer to the experimental data (see App. \ref{sec:A} for a comparison at all tested incident angles). 
\\
At 500 keV and $\theta_{0} = 0.33^{\circ}$ (Fig. \ref{fig:eff_500_1000}, left panel), MSC protons are also visible below a scattering angle of $4.5^{\circ}$, but with lower efficiencies at very small angles. The \textit{opt3} and \textit{opt4} lists result in the same distribution. SS efficiency rises at small angles for E$_{0}$ = 500 keV and is consistent among the entire range of experimental scattering angles.
\begin{figure}
\center
  \includegraphics[width=0.49\textwidth]{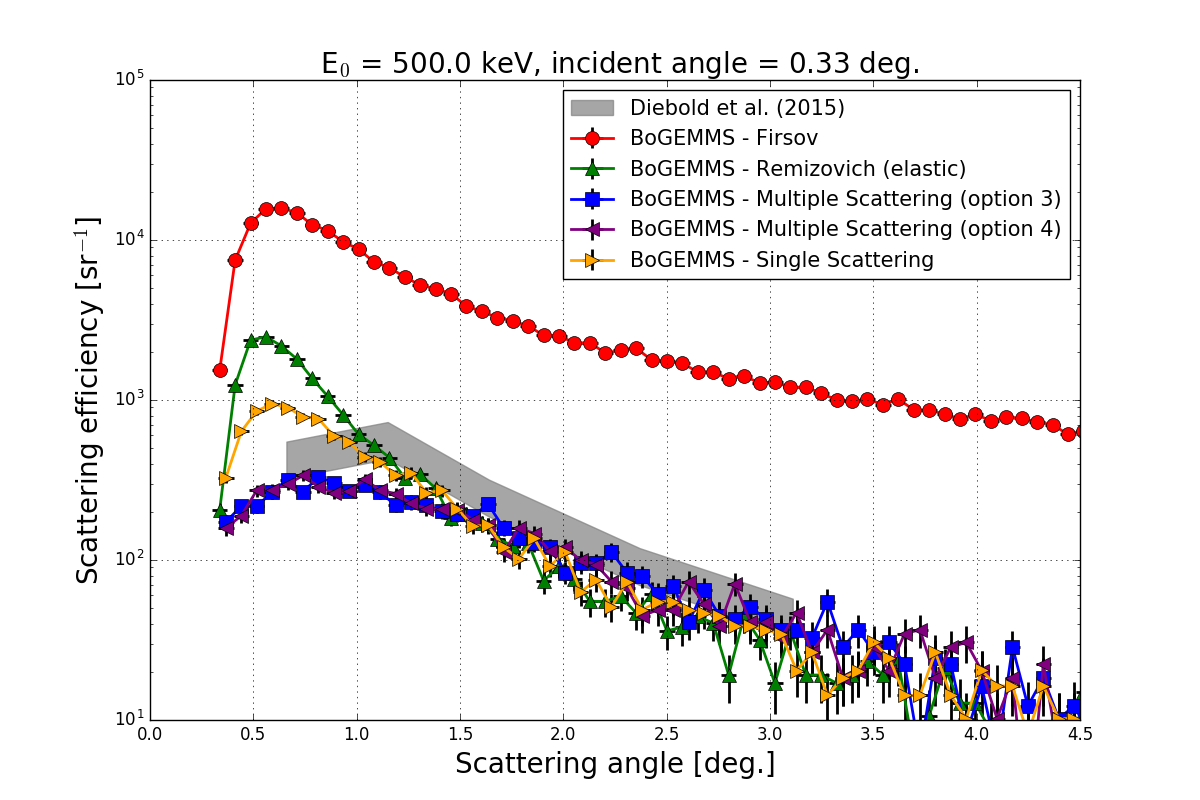}
    \includegraphics[width=0.49\textwidth]{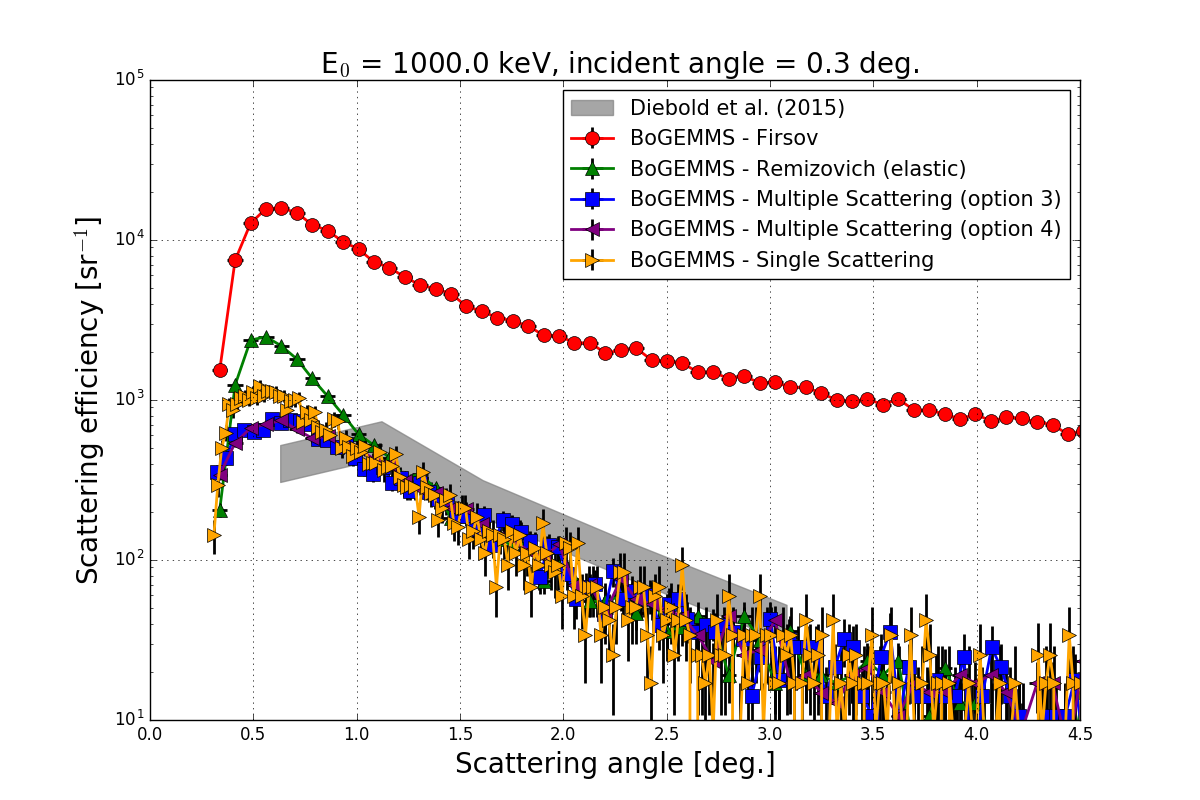}
\caption{\textit{Left panel:} Scattering efficiency at E$_{0}$ = 500 keV for an incident angle of $0.33^{\circ}$. \textit{Right panel:} Scattering efficiency at E$_{0}$ = 1000 keV for an incident angle of $0.3^{\circ}$.}
\label{fig:eff_500_1000}     
\end{figure}
If protons of 1000 keV are emitted (Fig. \ref{fig:eff_500_1000}, right panel), SS and MSC give similar results, with about a factor 2 of difference at the specular scattering angle. All models, except for the Firsov formula, generate consistent angular distributions for scattering angles higher than $1^{\circ}$.
Considering the three incident energies, 250, 500 and 1000 keV, the SS is the model that results in more similar results to the real data set. In Fig. \ref{fig:MSC} (right panel) we compare the SS induced proton angular distribution for two extreme energy values, 1000 and 50 keV. Above $\sim1.2^{\circ}$, for the present detection geometry, the proton energy has no effect on the scattering efficiency, while it becomes important at very small scattering angles, with about a factor 4 of difference between the two cases.

\begin{figure}
\center
  \includegraphics[width=0.49\textwidth]{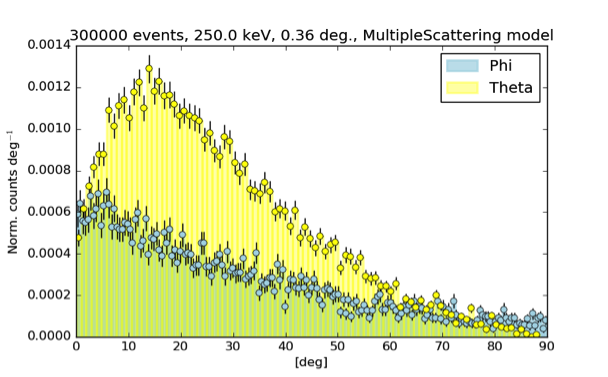}
  \includegraphics[width=0.49\textwidth]{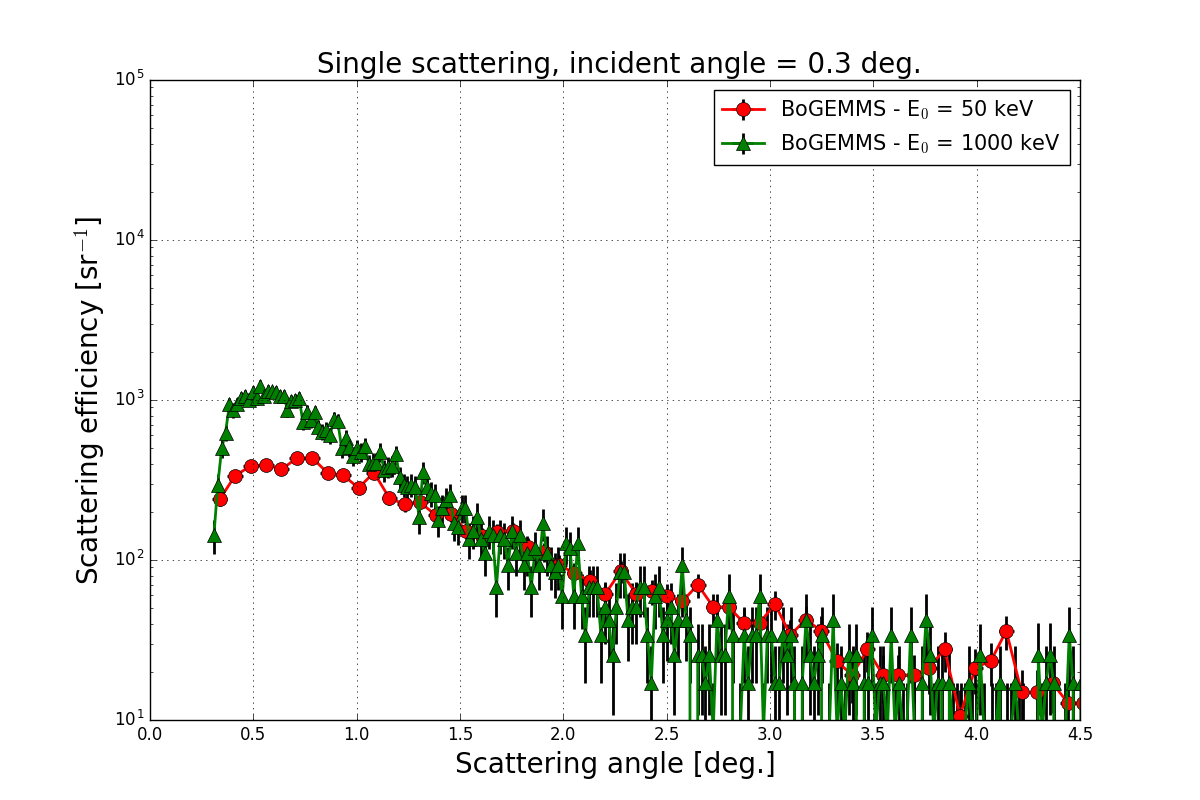}
\caption{\textit{Left panel:} The angular distribution in $\theta$ (yellow points) and $\phi$ (light blue points) of scattered protons if multiple scattering is used as physics interaction. \textit{Right panel:} Single scattering efficiency at E$_{0}$ = 50 and 1000 keV for an incident angle of $0.3^{\circ}$. }
\label{fig:MSC}     
\end{figure}
\subsection{Energy losses}

\begin{figure}
\center
  \includegraphics[width=0.7\textwidth]{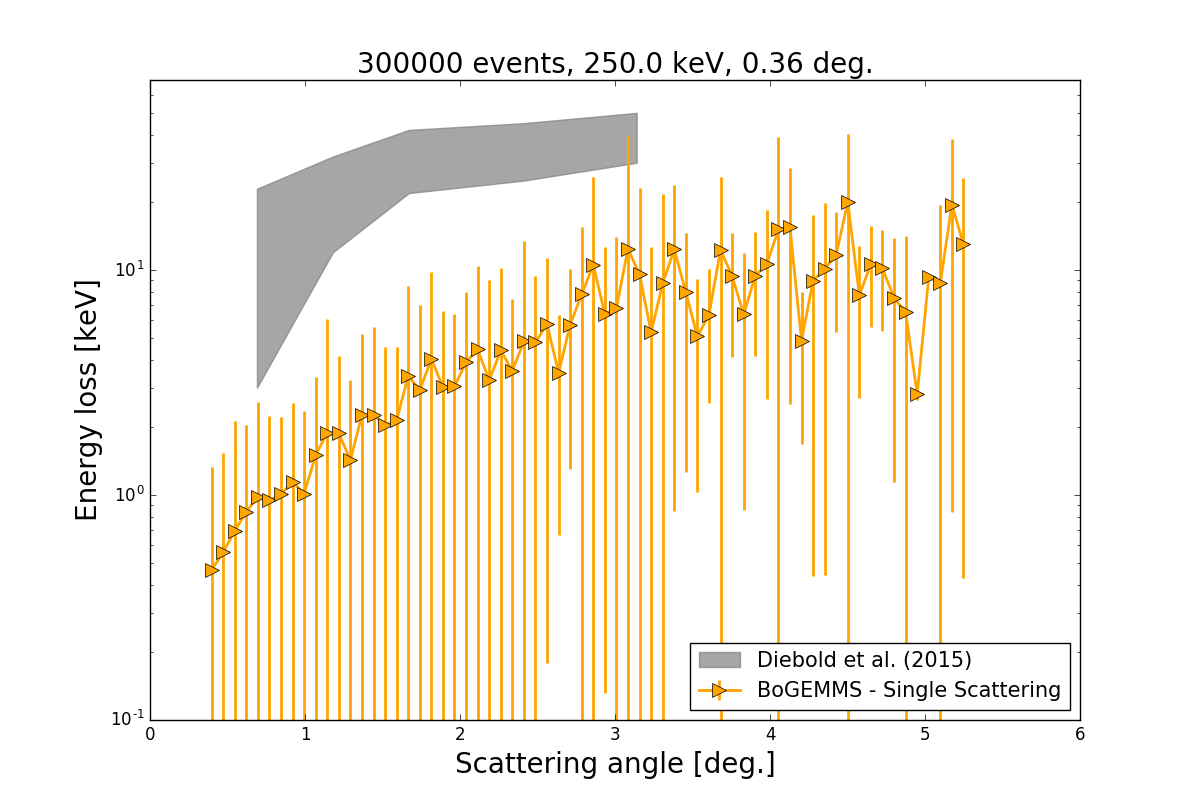}
\caption{The proton energy loss as a function of the scattering angle at E$_0$ = 250 keV for an incident angle of $0.36^{\circ}$. }
\label{fig:energy_250_1}     
\end{figure}

\begin{figure}
\center
  \includegraphics[width=0.49\textwidth]{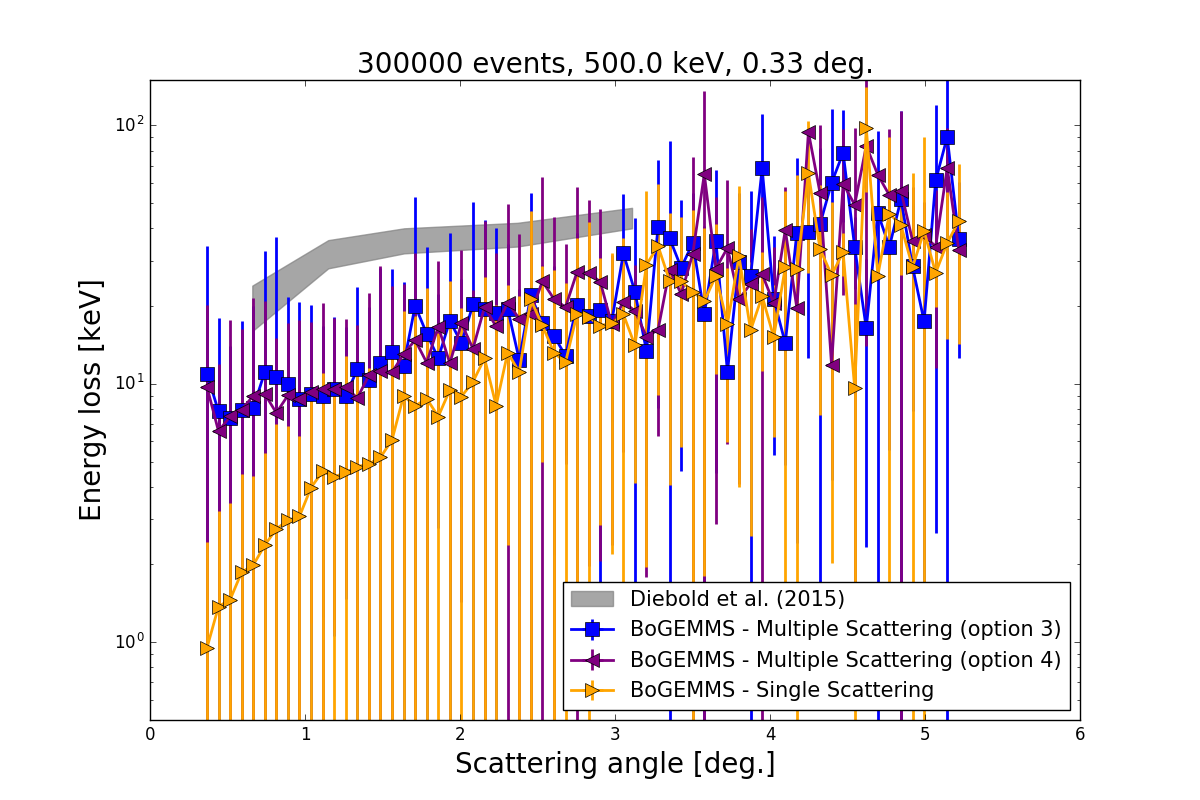}
    \includegraphics[width=0.49\textwidth]{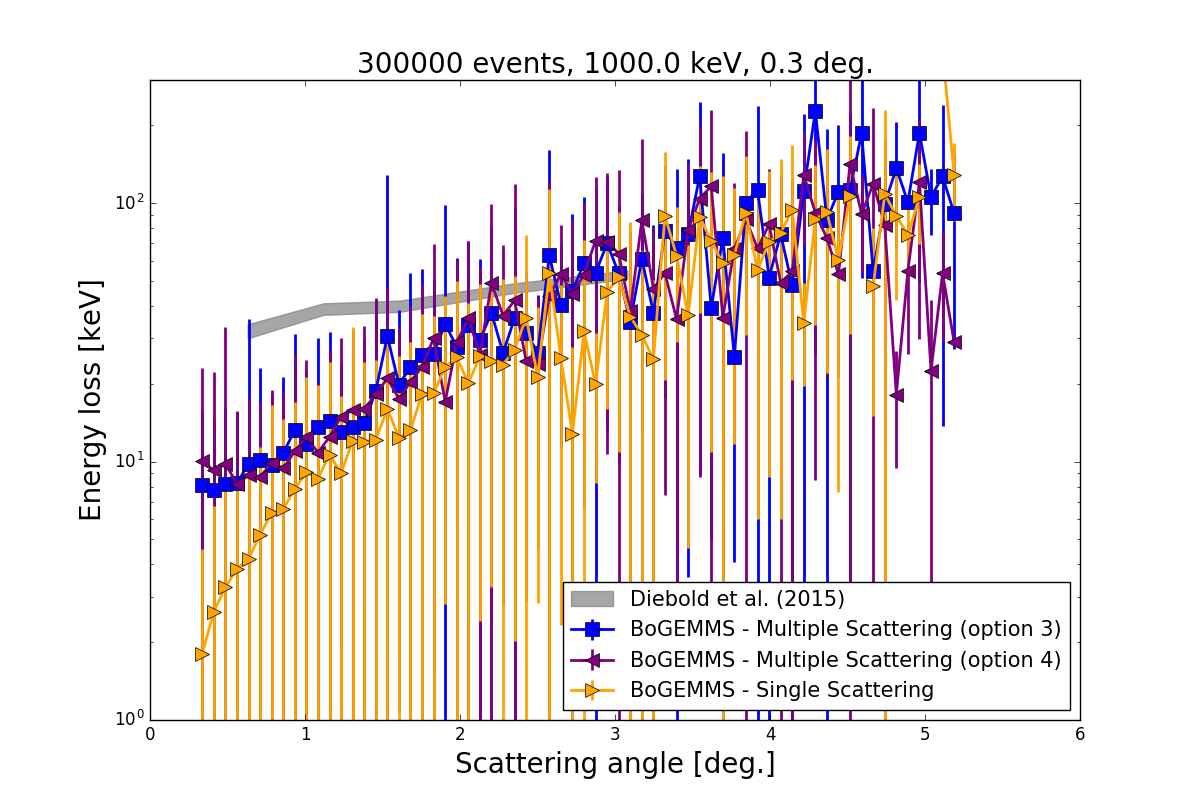}
\caption{\textit{Left panel:} The proton energy loss as a function of the scattering angle at E$_0$ = 500 keV for an incident angle of $0.33^\circ$ \textit{Right panel:} The proton energy loss as a function of the scattering angle at E$_0$ = 1000 keV for an incident angle of $0.3^\circ$ deg.}
\label{fig:energy_500_1000}     
\end{figure}

The measured energy loss is obtained by Gaussian fits of the energy distribution of both the incident and scattered protons. In our case the statistics for each bin of the scattering angle is not enough to produce a fit and the energy loss is given by the mean of the proton energies, for each bin, subtracted by the incident energy (error bars are the standard deviation of the energy distribution). The experimental results, that range from $\langle \rm E_{\rm Loss}\rangle = 13$ keV ($\theta_{0} = 0.69^{\circ}$) at E$_{0}$ = 250 keV to 54 keV ($\theta_{0} = 3.09^{\circ}$) at E$_{0}$ = 1000 keV, give higher energy losses than the few keVs expected from \cite{1987PhRvB..36....7K}. From these findings, the percentage of energy lost in each scattering seems to be constant, with a value of $\sim5\%$ with respect to the initial energy. Remizovich and coauthors \cite{1980JETP...52..225R} find a constant behaviour in the percentage of energy lost in the scattering similar to what obtained in the present measurements.
%The same behaviour is found by Remizovich and coauthors \cite{1980JETP...52..225R}, with percentage values consistent within a factor 2-4.  
Since the Remizovich model is implemented in the elastic approximation, we only compare the simulation using the inelastic SS and MSC scattering interactions. 
\\
At 250 keV and for $\theta_0 = 0.36^{\circ}$ (see Fig. \ref{fig:energy_250_1}), SS gives energy losses less than 1 keV at the specular reflection angles, and up to $\sim10$ keV for larger scattering angles, about 10 times less than the experimental data. The same behavior is obtained at higher incident angles (see Fig. \ref{fig:energy_250} of App. \ref{sec:B}). At 500 and 1000 keV (see Fig. \ref{fig:energy_500_1000}), SS and MSC give similar results at large scattering angle, as seen for the scattering efficiency, but in the specular range the multiple scattering induces higher, $>10$ keV, energy losses, close to the real data points. No difference is observed between the electromagnetic \textit{opt3} and \textit{opt4}.
For larger incident angles (see Figure 72 and Figure 73 in Appendix) the multiple scattering results in energy losses well consistent with the observed ones if the proton energy is in the 500-1000 keV band.

%\subsection{Geant4 vs Ray-tracing}\label{sec:ray}
%The Remizovich implementation in the Geant4 code has been compared to the ray-tracing simulation obtained by the work of [REF] for an independent verification. The two curves, as shown in Figure 15, are consistent.

%\begin{figure}
%\center
%  \includegraphics[width=0.6\textwidth]{ray_g4.png}
%\caption{The scattering efficiency induced by the Remizovich model at E0 = 1000 keV for an incident angle of $0.3^{\circ}$. The red and blue lines refer to the Geant4 and ray-tracing simulation respectively, the green points are the laboratory measurements.}
%\label{fig:ray}     
%\end{figure}
\subsection{Code optimization issues}
In the present Geant4 implementation, the Remizovich probability distribution, which depends on the incident proton angle, is computed at each interaction. This approach is feasible for the Firsov model, where only the polar angle is randomly generated, but not for the Remizovich model, that is a bivariate distribution in $\phi$ and $\theta$, with a CPU processing time higher than a factor 100 with respect to the integrated Firsov version at the highest energies. Table \ref{tab:opt} compares the simulation CPU time for all the tested models in units of time required by multiple scattering (\textit{emstandard\_option3}). The performance test refers to a run of $10^{4}$ protons at 250 keV for an incident angle of 0.36 degrees.
\\
%
% For tables use
\begin{table}
\center
% table caption is above the table
\caption{CPU time, normalized to the multiple scattering (opt3) time, required to run the simulation for the tested physics models.}
\label{tab:opt}       % Give a unique label
% For LaTeX tables use
\begin{tabular}{clclclc|c|c}
\hline\noalign{\smallskip}
& MSC (\textit{opt3}) & MSC (\textit{opt4}) & SS & Remizovich & Firsov  \\
\noalign{\smallskip}\hline\noalign{\smallskip}
CPU time & 1 & 1.1 & 21.3 & 366.7 & 3.1 \\
\noalign{\smallskip}\hline
\end{tabular}
\end{table}
Thanks to the very simple mass model used in the proton scattering test, we were able to perform the present simulations by a dedicated fine-tuning of the angle resolution (see Sec. \ref{sec:firsov}) and the optimization of the Geant4 algorithm. The latter produced a gain of $\sim50\%$ in CPU time. 
In case of future releases of the code for its use in simulation campaigns of the background of X-ray space telescopes, as the case of ATHENA, where complex geometries and high statistics are required, the optimization of the Remizovich Geant4 implementation is mandatory. Besides the use of the Geant4 multi-threading option already built in the standard Geant4 10.2 release, possible solutions could be the use of for loop parallelization APIs (e.g. openMP\footnote{http://openmp.org/wp/}) or loading the probability distribution from an external physics database.

\section{Summary}
The most recent laboratory measurements of \cite{2015ExA....39..343D}, testing the angular and energy distribution of protons scattering by X-ray mirror shells at glancing angles, are the only available data set for the physics assessment of the interaction behind the soft proton scattering in X-ray space telescopes. After the implementation and verification of two new physics models (the elastic Remizovich model and the azimuthally focused Firsov model), we simulate the experimental set-up to find out which model, among the new solutions and the ones already built in the standard Geant4 library (single and multiple scattering), better reproduce the observation. 
Our results can be summarized as follows:
\begin{itemize}
\item if we consider the proton distribution at 250 keV, Remizovich and SS are well consistent with the experimental scattering efficiency except for very small, $<1^{\circ}$, scattering angles, where higher efficiencies are found in the simulation;
\item the SS induced scattering efficiency at small angles is the closest to the observation, but the energy losses are a factor 10 less than the experimental ones;
\item MSC can not reproduce the angular distribution of protons at 250 keV, and it is not feasible to simulate soft proton funnelling by X-ray optics;
\item at large scattering angles, the SS and Remizovich solutions give consistent scattering efficiencies;
\item because of the large spread in the azimuthal scattering angle, the use of the Firsov model with $\phi=0$ overestimates more than 10 times the real distribution;
\item no differences are found between the electromagnetic \textit{opt3} and \textit{opt4} list MSC settings.
\end{itemize}

\section{Conclusions and remarks}
%Remizovich formula for inelastic scattering foresees energy losses close, within a factor 2-4, to what obtained in the latest measurements. Its implementation in Geant4 should provide a better description of the experimental results. 
Since the Remizovich model in its approximated form has proven a general consistency with the measurements, the present Remizovich Geant4 implementation will not only be included in the \textit{Space Physics list} of the ATHENA Radiation Environment Models and X-Ray Background Effects Simulators (AREMBES), but it is currently being used as basis for the development of the official Geant4 Remizovich classes to be included in the next release of the Geant4 toolkit.
This activity is carried on under the responsibility of Geant4 collaboration members within the \textit{Low Energy working group} \footnote{https://twiki.cern.ch/twiki/bin/view/Geant4/LowEnergyElectromagneticPhysicsWorkingGroup}.
The implementation in Geant4 of the Remizovich formula for inelastic scattering, predicting energy losses close, within a factor 2-4, to what obtained in the latest measurements, should provide a better description of the experimental results and it is planned in the future activities. 
\\
Despite the obtained results represent a first step toward the development of a Geant4 grazing angle low energy proton scattering model dedicated to X-ray optics on board space missions, it must be pointed out that the lowest proton energy obtained in this experimental facility, 250 keV, is above the range of interest of the protons that induce background in X-ray missions like Chandra and XMM-Newton, where protons below 100  keV are the ones that mostly deposit energies inside the detectors sensitivity band, inducing a poorly reproducible background component.
\\
Because of (i) the lack of data at very low scattering angles, (ii) the lower energy losses obtained with respect to the experiment, (iii) and the high energy of the incoming protons, it is not yet possible to advertise any of the tested models as the most accurate one for the simulation of the scattering of low energy protons experienced by ATHENA X-ray optics. 
However, further developments are both ongoing on the software side and studies planned as new data will be available in the near future, focused at lower proton energies and lower scattering angles in order to definitively select one of those models to represent the proton scattering behavior in this regime.

\begin{acknowledgements}
This work is supported by the ESA Contract No 4000116655/16/NL/BW. The AHEAD project (grant agreement n. 654215) which is part of the EU-H2020 programm is acknowledged for partial support.
\end{acknowledgements}

% BibTeX users please use one of
%\bibliographystyle{spbasic}      % basic style, author-year citations
%\bibliographystyle{spmpsci}      % mathematics and physical sciences

\bibliographystyle{spphys}       % APS-like style for physics
\bibliography{fioretti_bib}   % name your BibTeX data base
\newpage 
\appendix
\section{Scattering efficiency}\label{sec:A}

\begin{figure}[h!]
\center
  \includegraphics[width=0.44\textwidth]{eff_250_1_label.png}
  \includegraphics[width=0.44\textwidth]{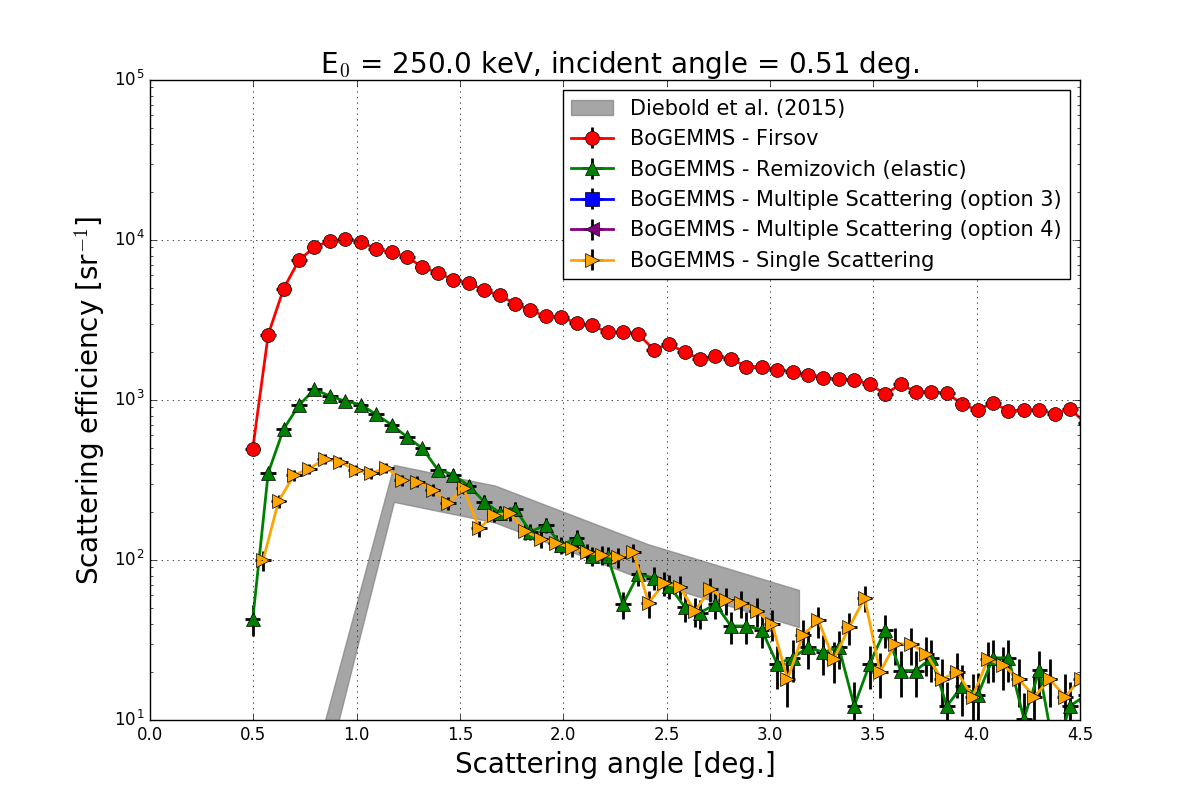}\\
  \includegraphics[width=0.44\textwidth]{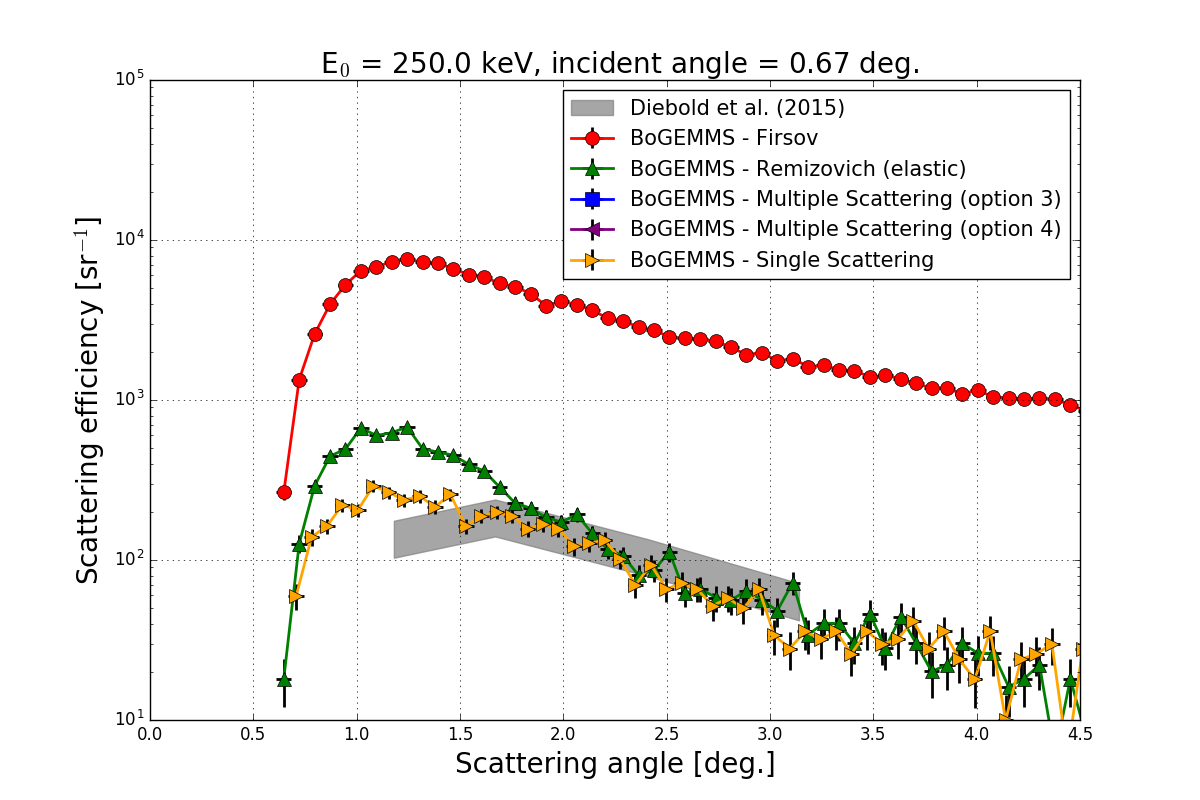}
  \includegraphics[width=0.44\textwidth]{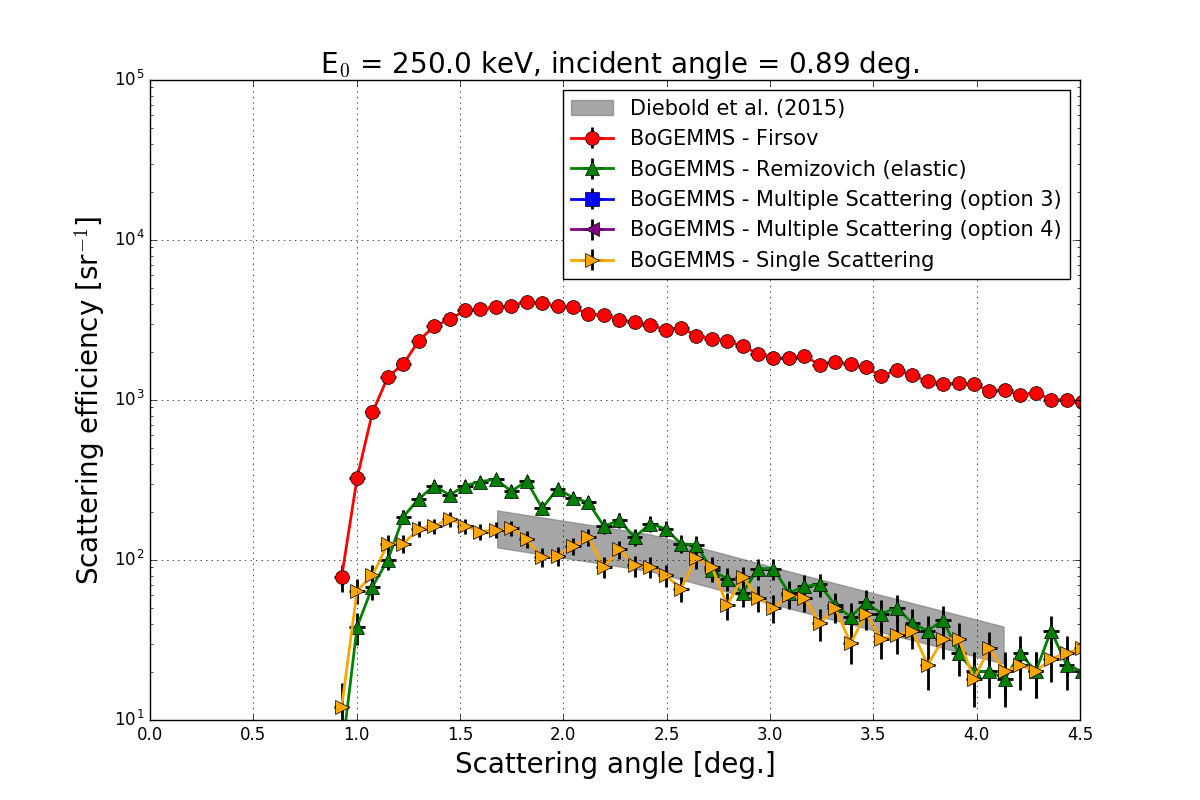}\\
    \includegraphics[width=0.44\textwidth]{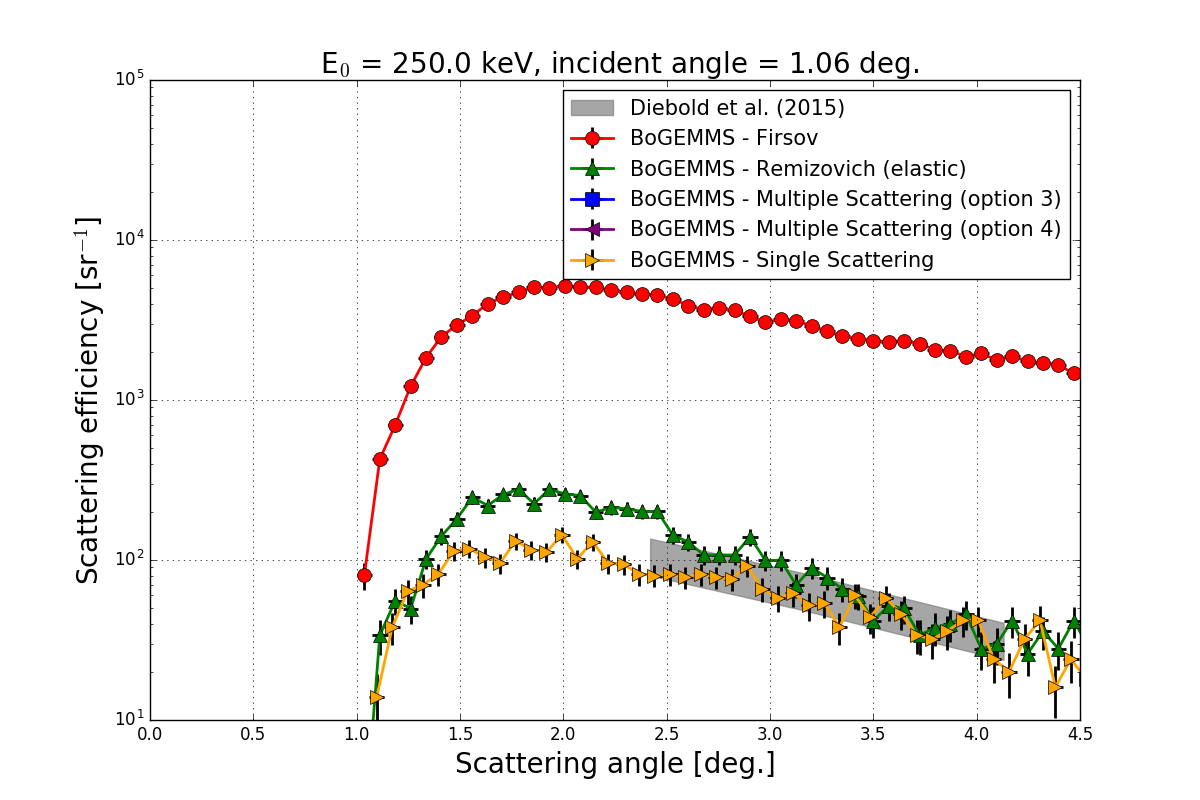}
  \includegraphics[width=0.44\textwidth]{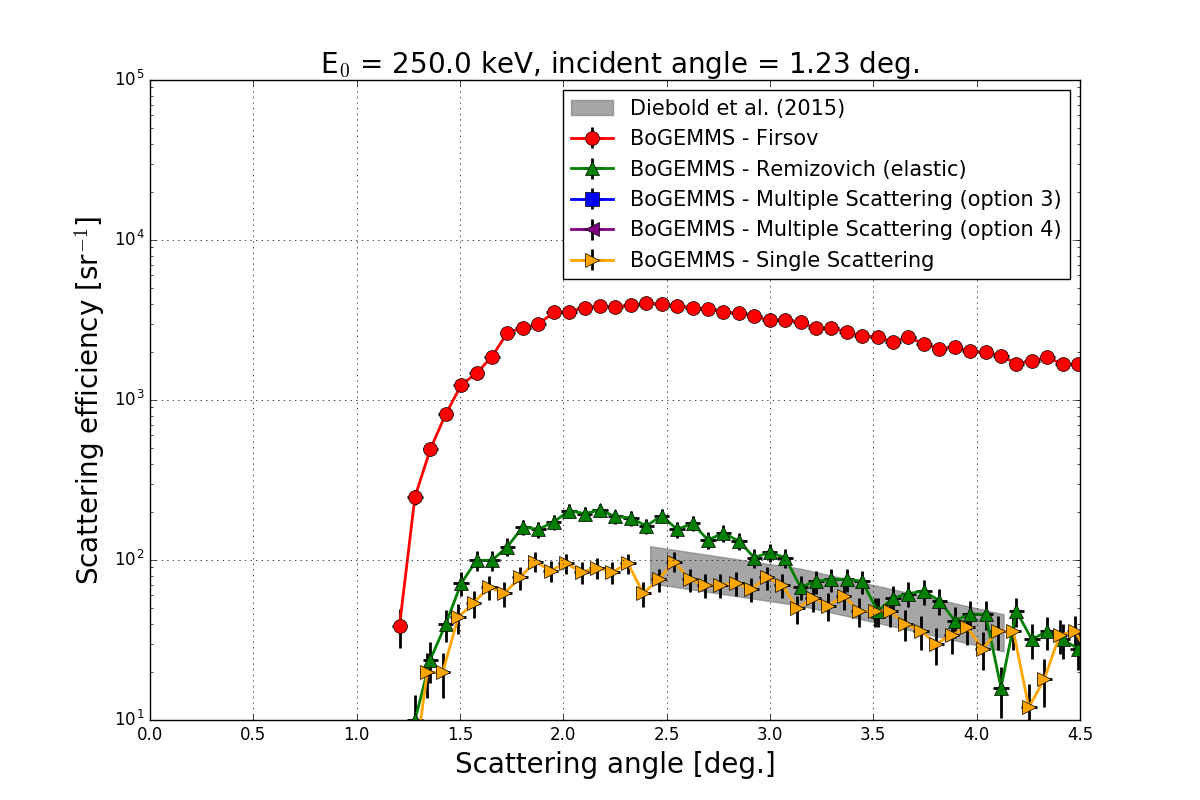}
\caption{Scattering efficiency at E$_{0}$ = 250 keV for an incident angle ranging from $0.36^{\circ}$ to $1.23^{\circ}$.}
\label{fig:eff_250}     
\end{figure}
\newpage
\begin{figure}[h!]
\center
  \includegraphics[width=0.49\textwidth]{eff_500_1_label.png}
  \includegraphics[width=0.49\textwidth]{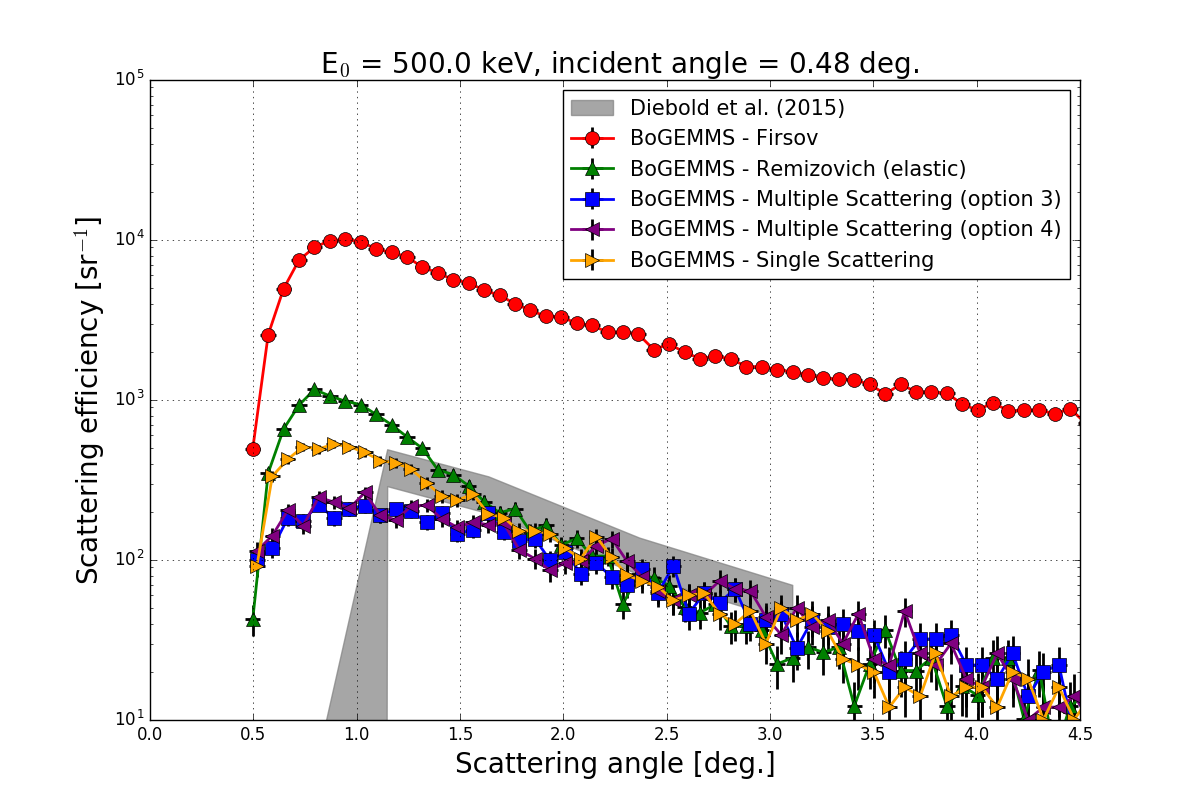}\\
  \includegraphics[width=0.49\textwidth]{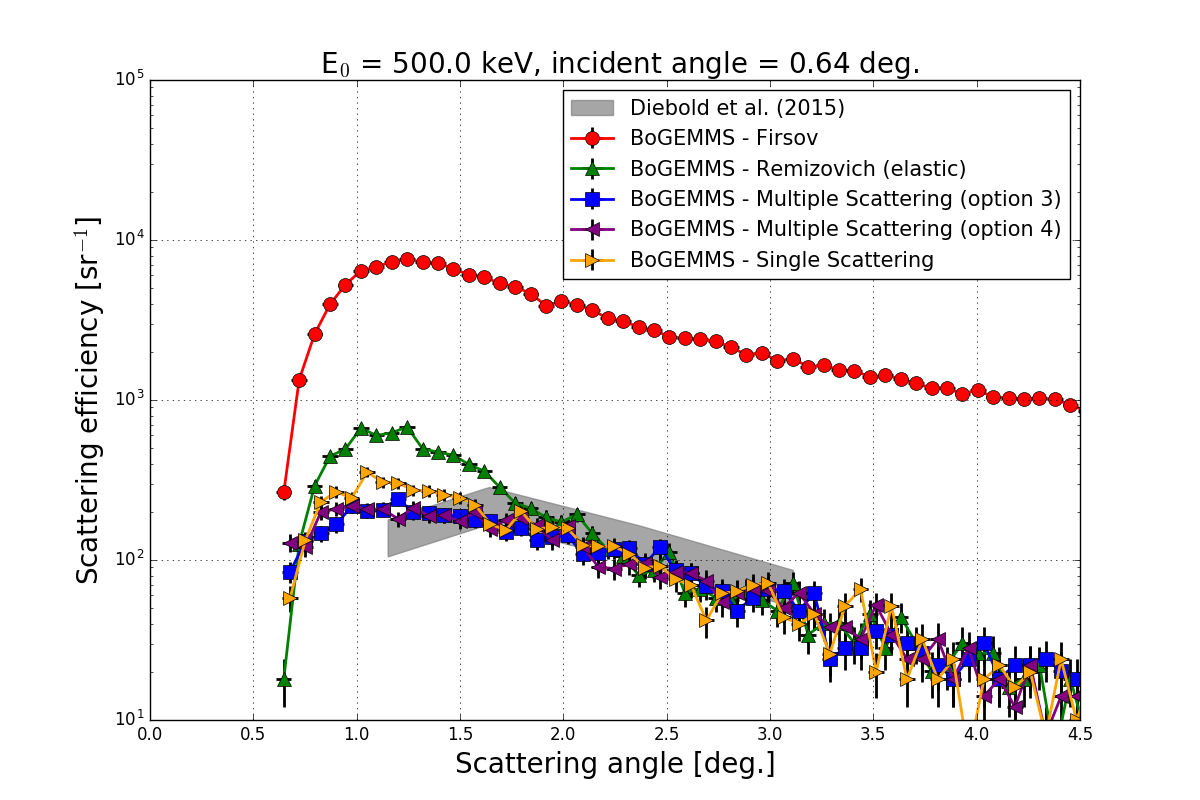}
  \includegraphics[width=0.49\textwidth]{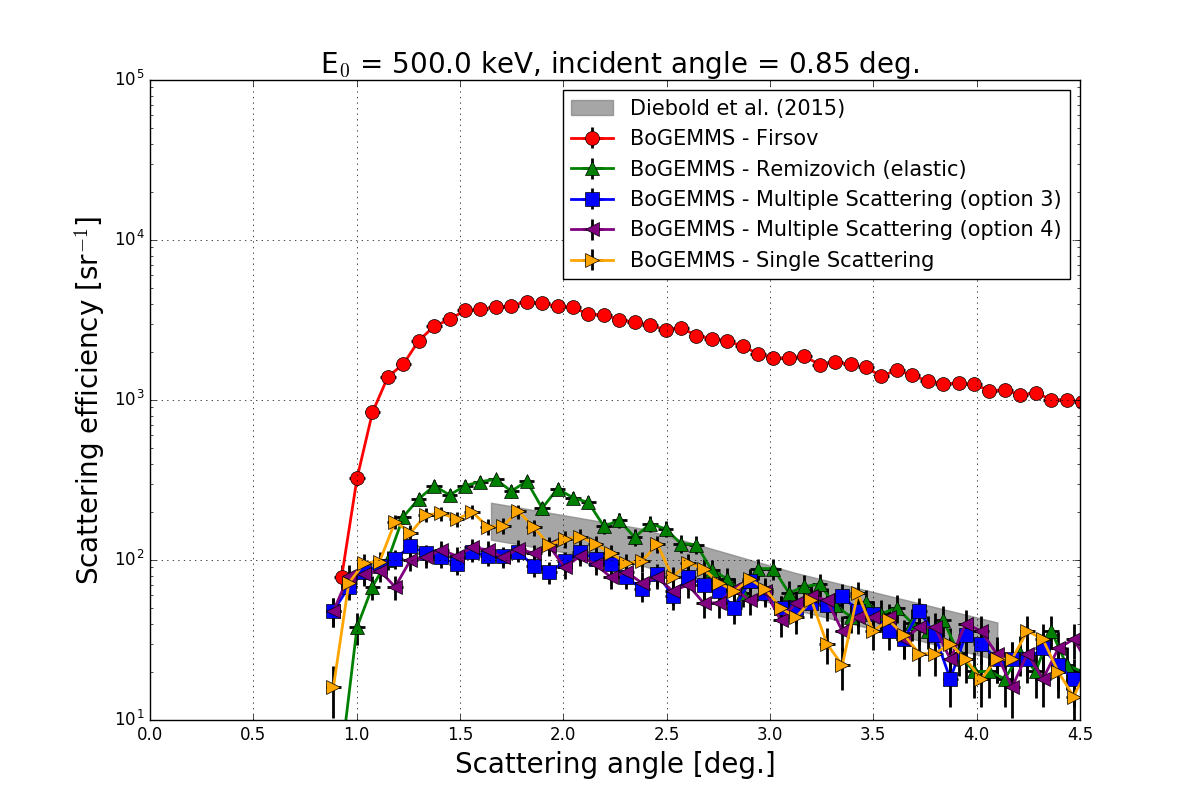}\\
    \includegraphics[width=0.49\textwidth]{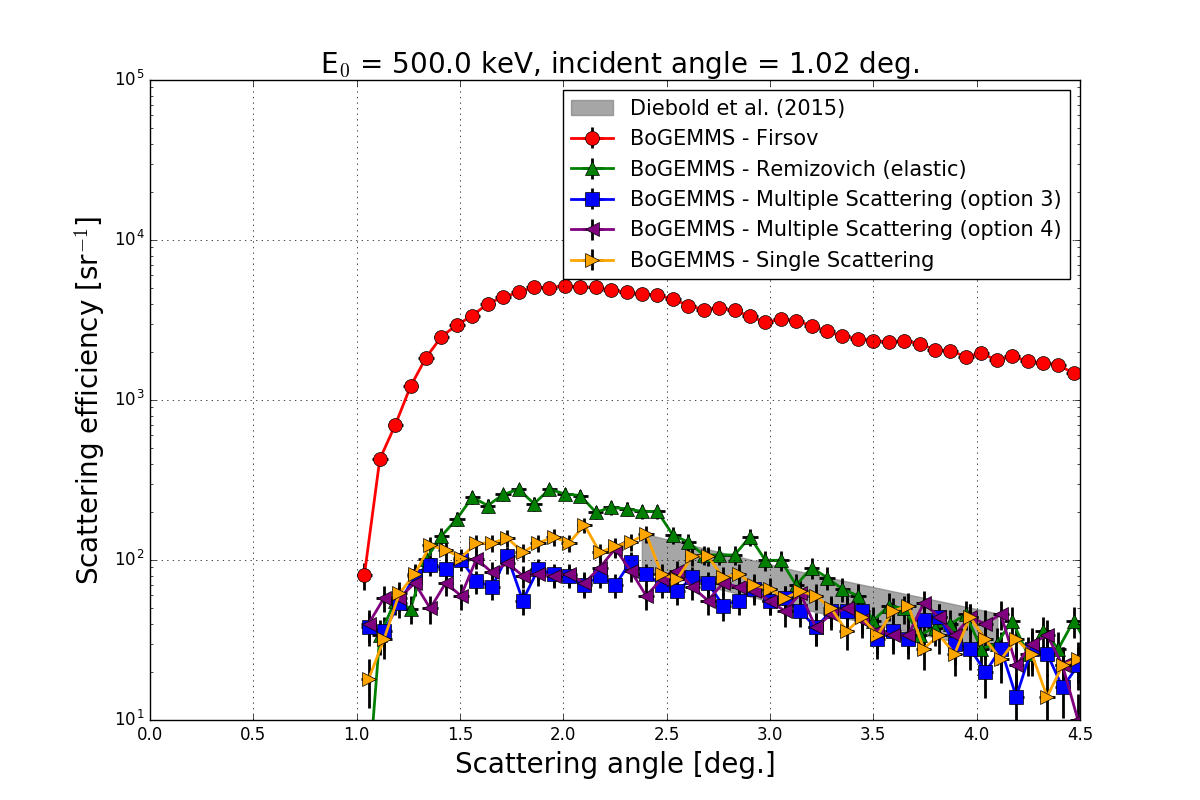}
  \includegraphics[width=0.49\textwidth]{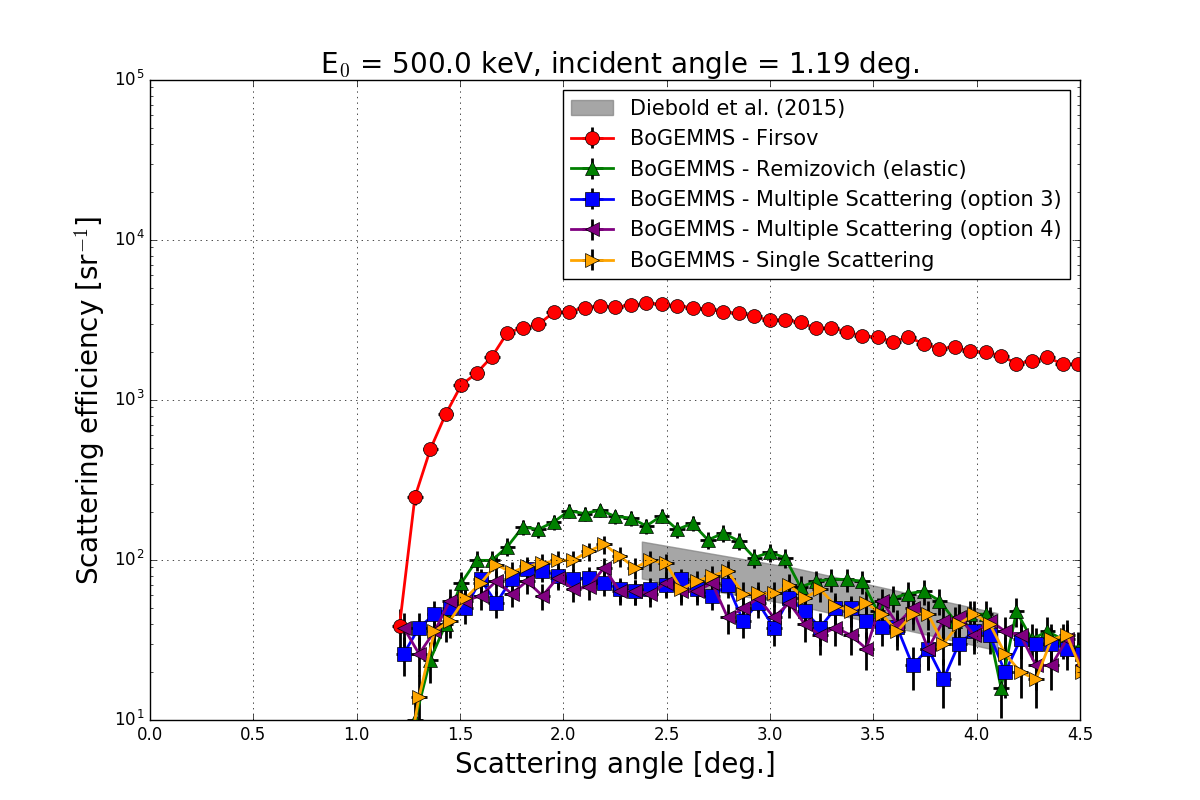}
\caption{Scattering efficiency at E$_{0}$ = 500 keV for an incident angle ranging from $0.33^{\circ}$ to $1.19^{\circ}$.}
\label{fig:eff_500}     
\end{figure}
\newpage
\begin{figure}[h!]
\center
  \includegraphics[width=0.49\textwidth]{eff_1000_1_label.png}
  \includegraphics[width=0.49\textwidth]{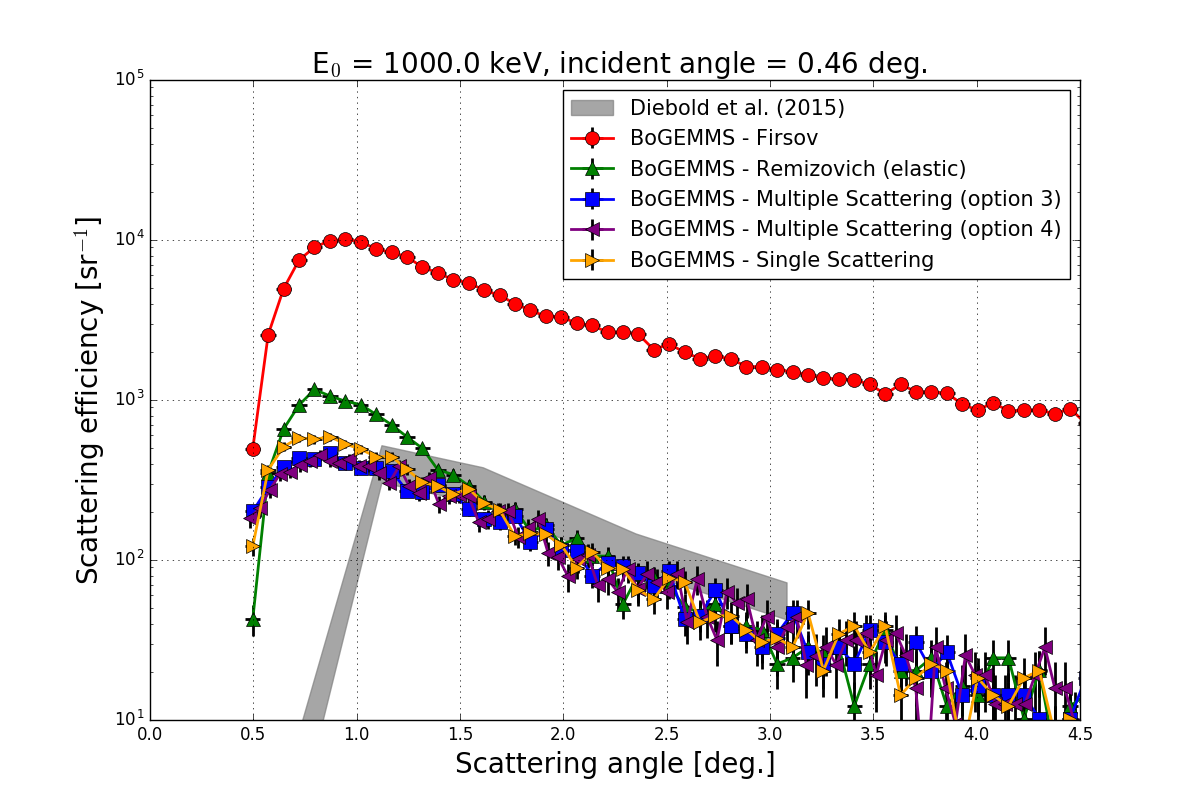}\\
  \includegraphics[width=0.49\textwidth]{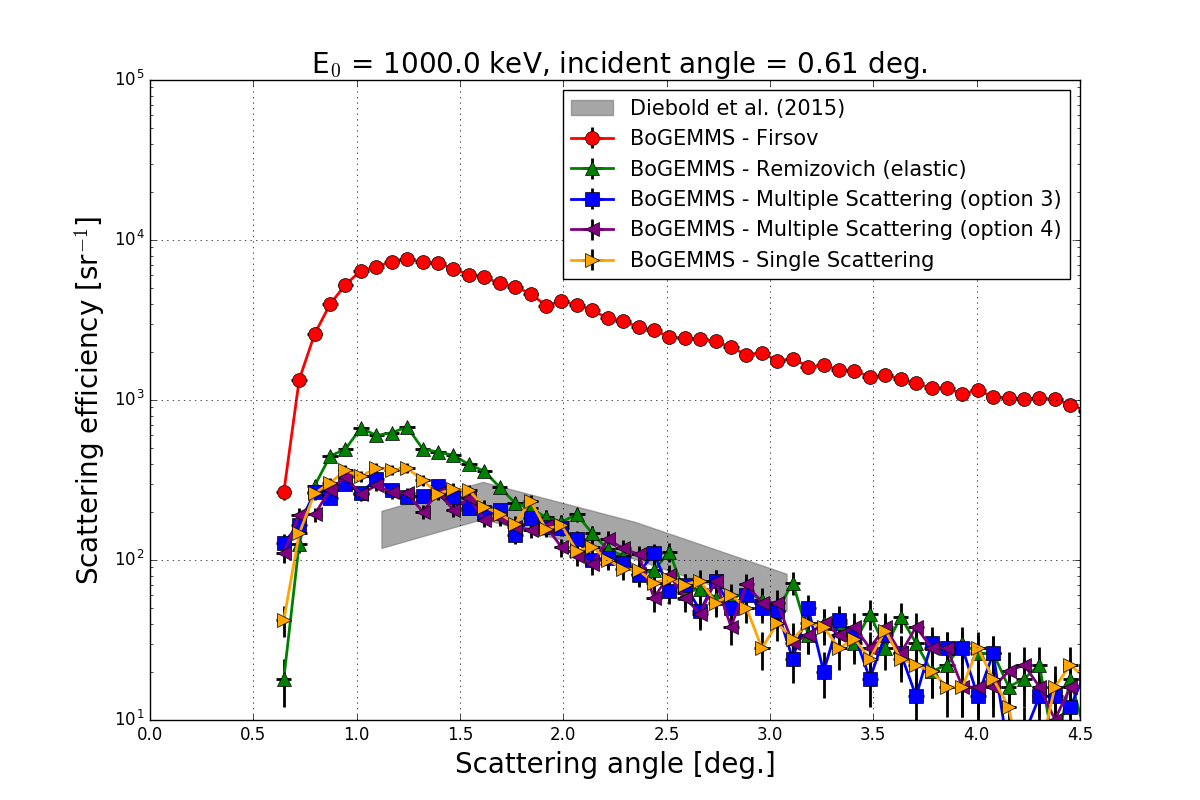}
  \includegraphics[width=0.49\textwidth]{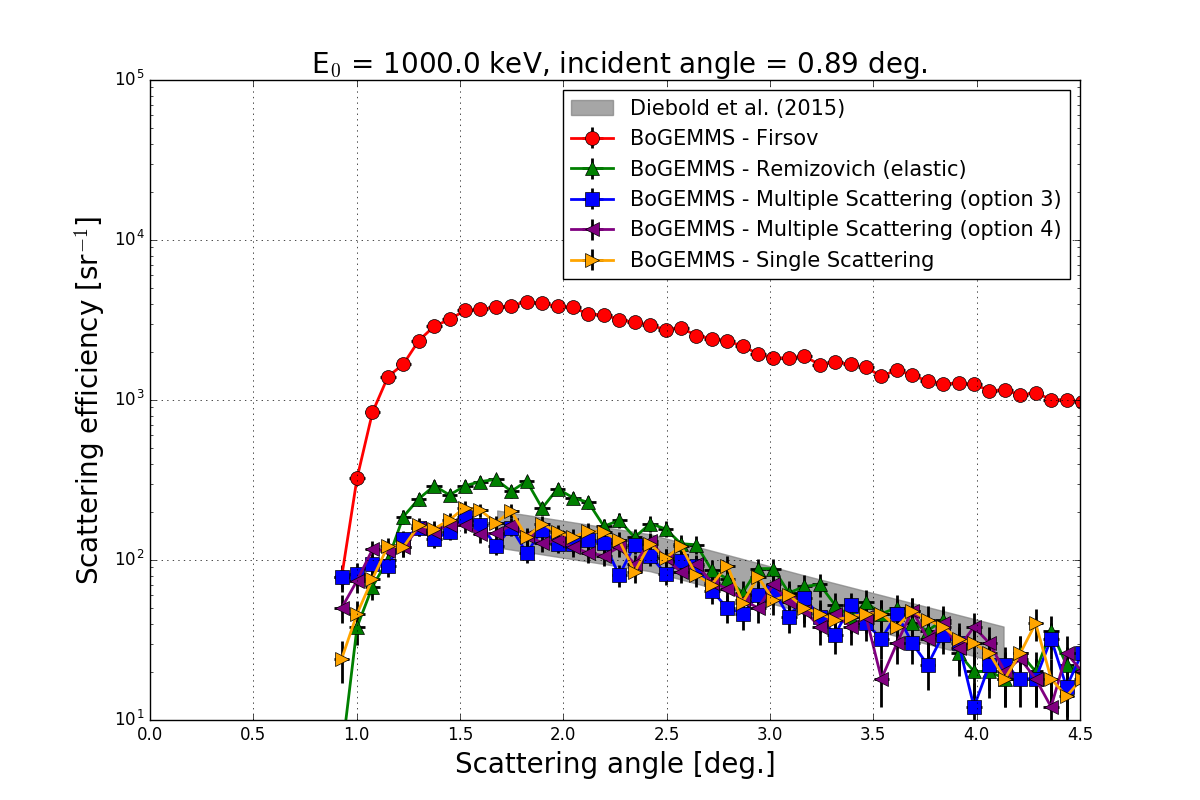}\\
    \includegraphics[width=0.49\textwidth]{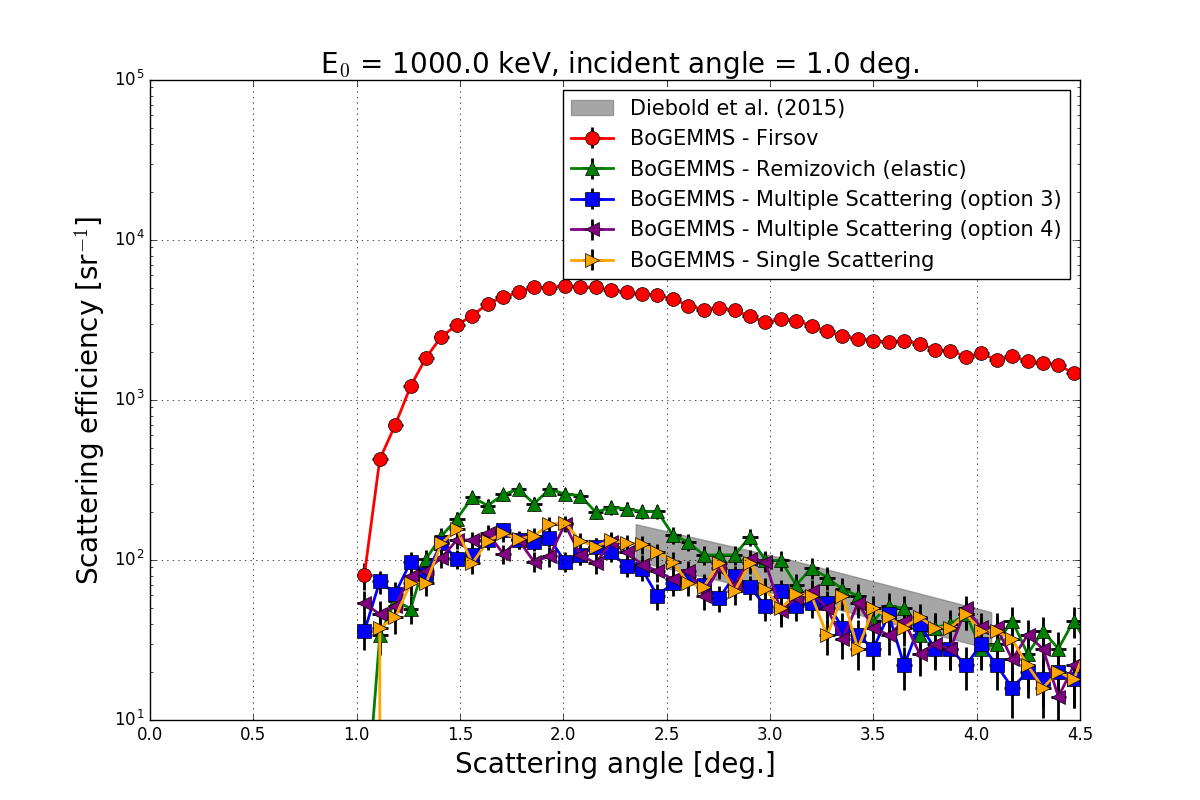}
  \includegraphics[width=0.49\textwidth]{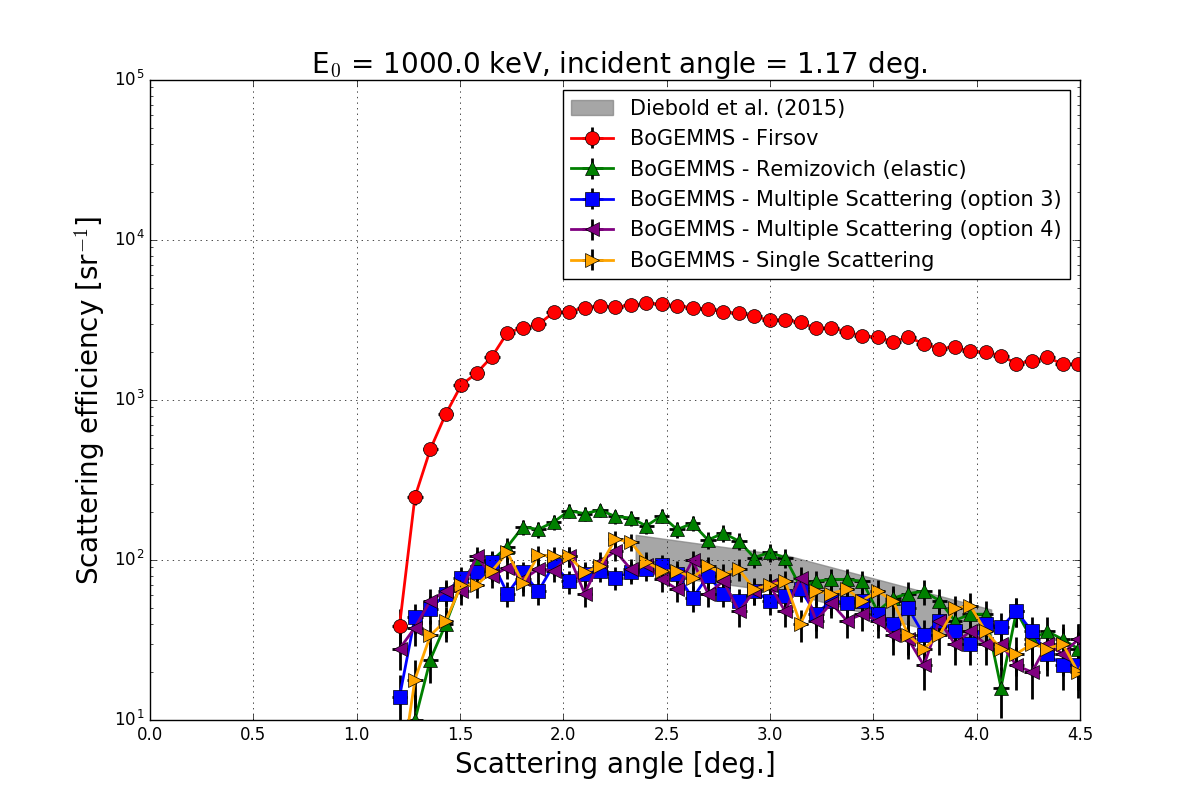}
\caption{Scattering efficiency at E$_{0}$ = 1000 keV for an incident angle ranging from $0.3^{\circ}$ to $1.17^{\circ}$.}
\label{fig:eff_1000}     
\end{figure}

\newpage
\section{Energy losses}\label{sec:B}
\begin{figure}[h!]
\center
  \includegraphics[width=0.49\textwidth]{energy_250_1_label.png}
  \includegraphics[width=0.49\textwidth]{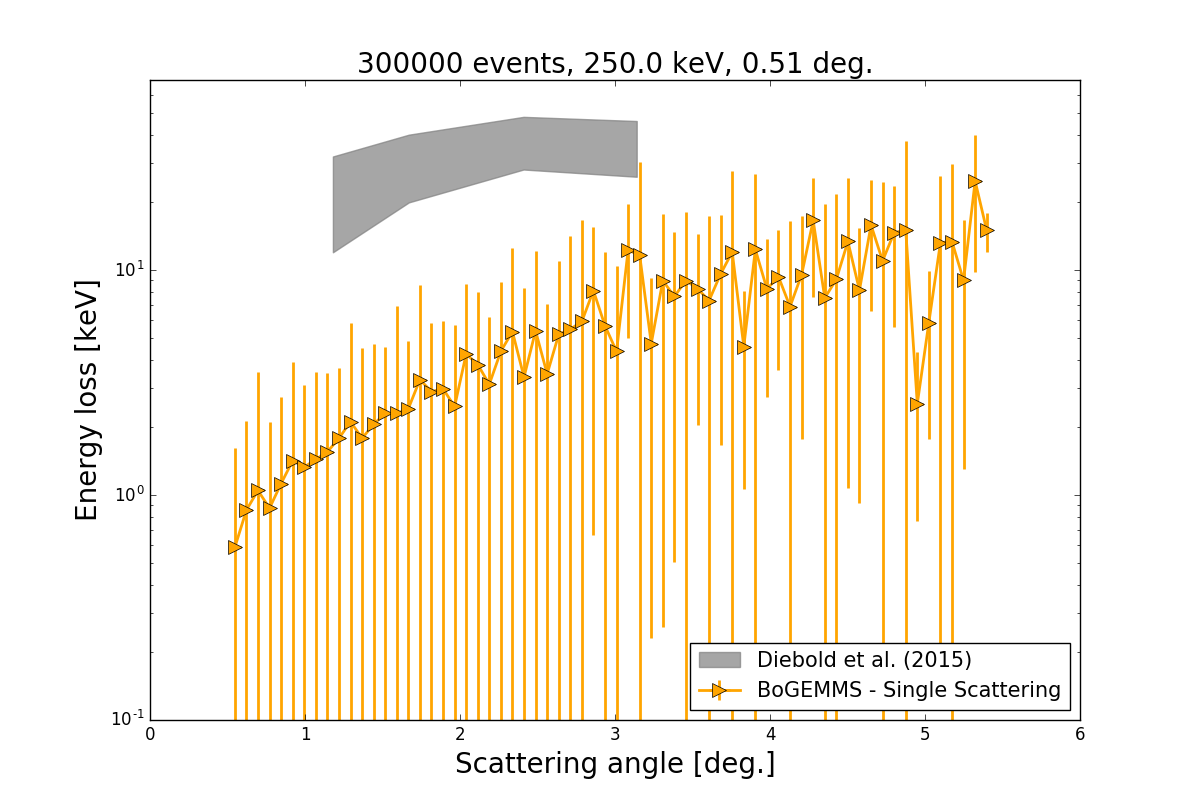}\\
  \includegraphics[width=0.49\textwidth]{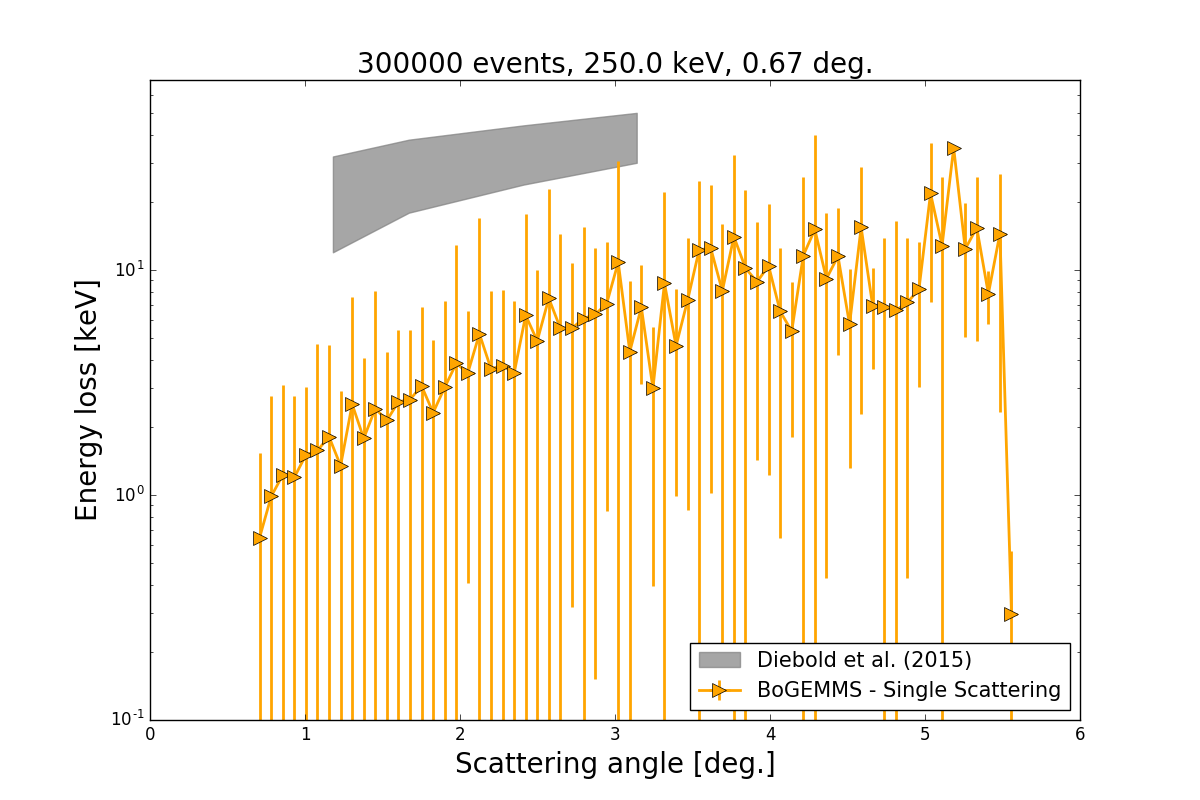}
  \includegraphics[width=0.49\textwidth]{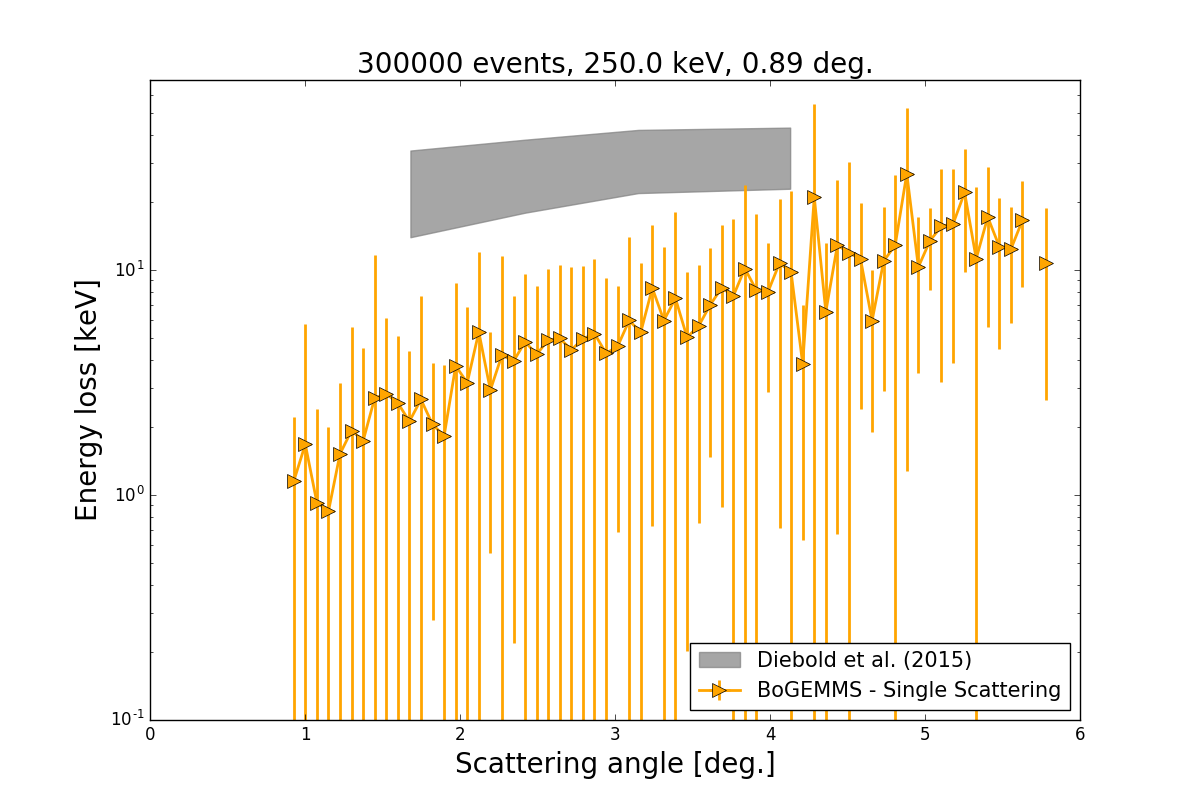}\\
    \includegraphics[width=0.49\textwidth]{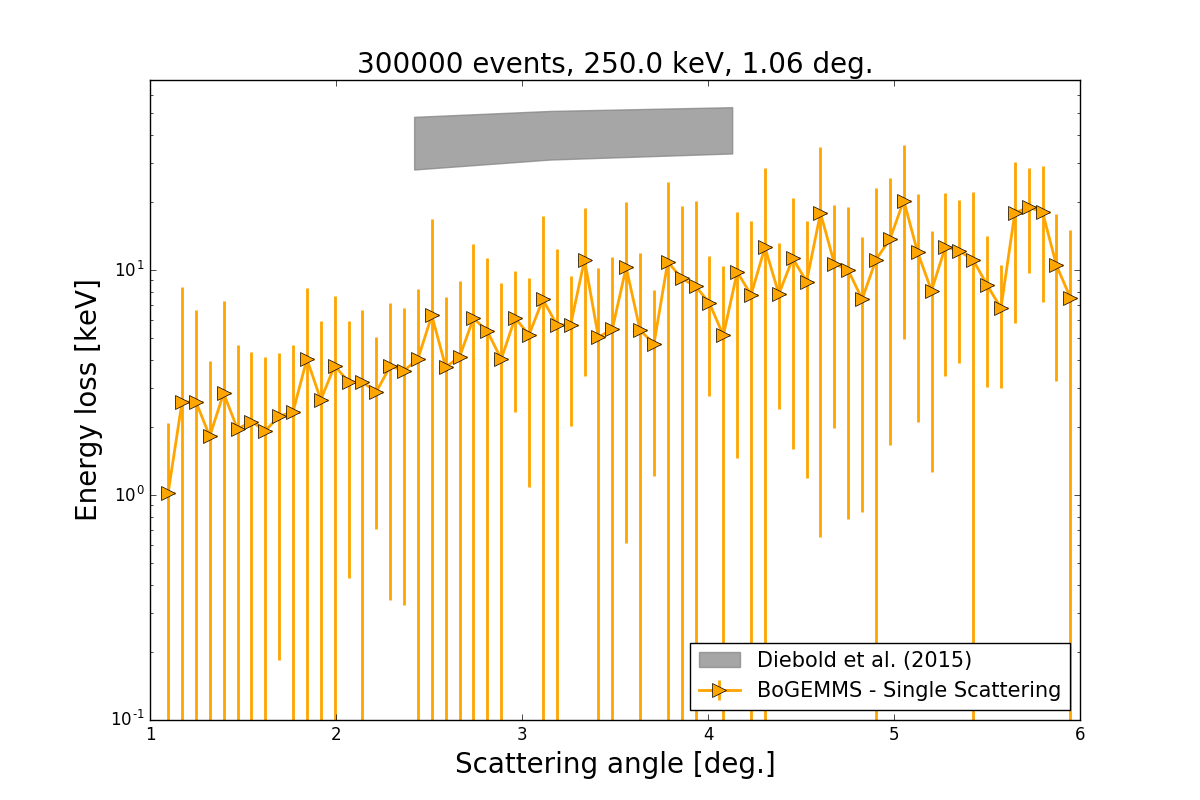}
  \includegraphics[width=0.49\textwidth]{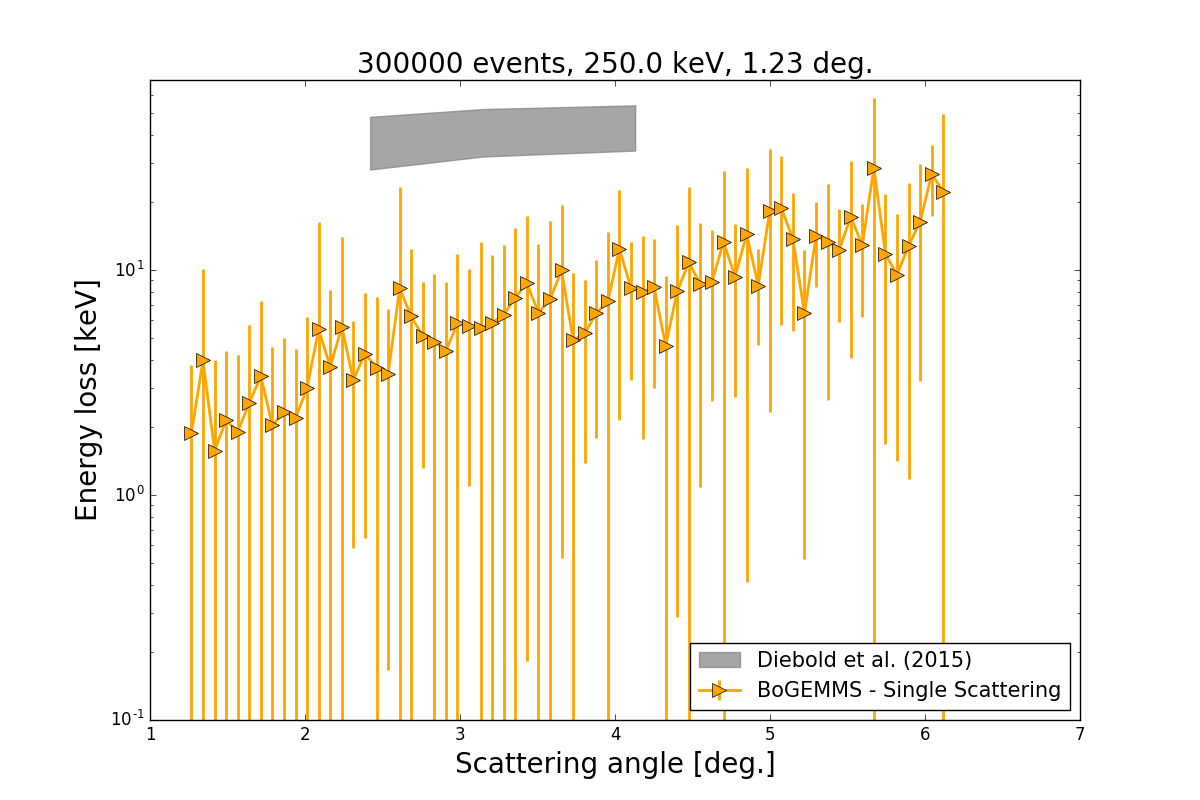}
\caption{The proton energy loss as a function of the scattering angle at E$_0$ = 250 keV in the $0.36^\circ - 1.23^\circ$ incident angle range. }
\label{fig:energy_250}     
\end{figure}
\begin{figure}[h!]
\center
  \includegraphics[width=0.49\textwidth]{energy_500_1_label.png}
  \includegraphics[width=0.49\textwidth]{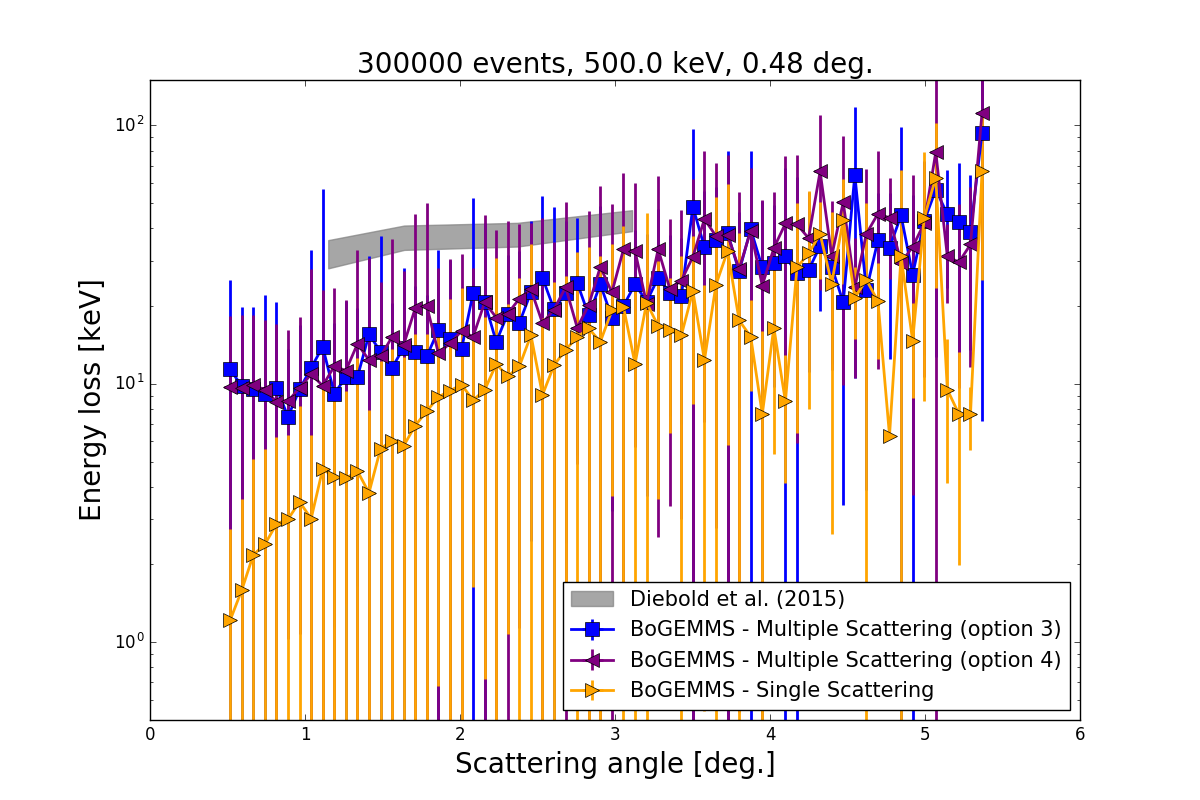}\\
  \includegraphics[width=0.49\textwidth]{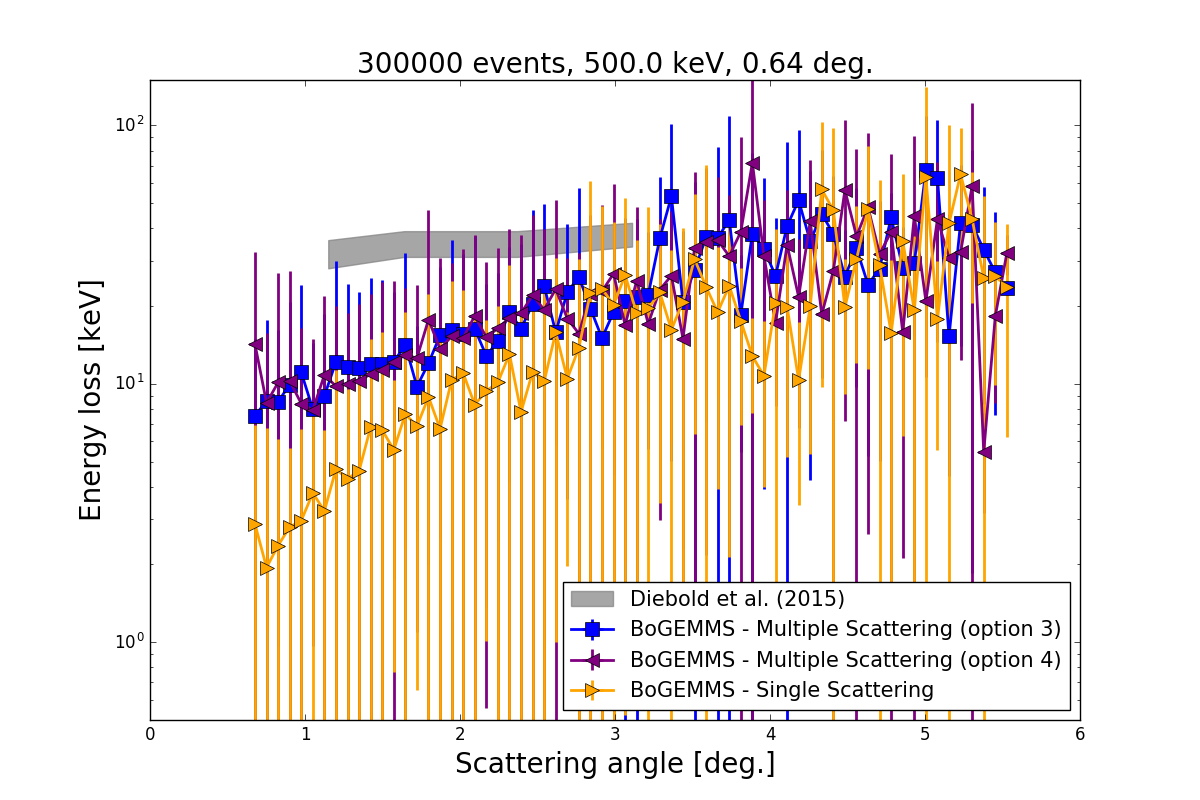}
  \includegraphics[width=0.49\textwidth]{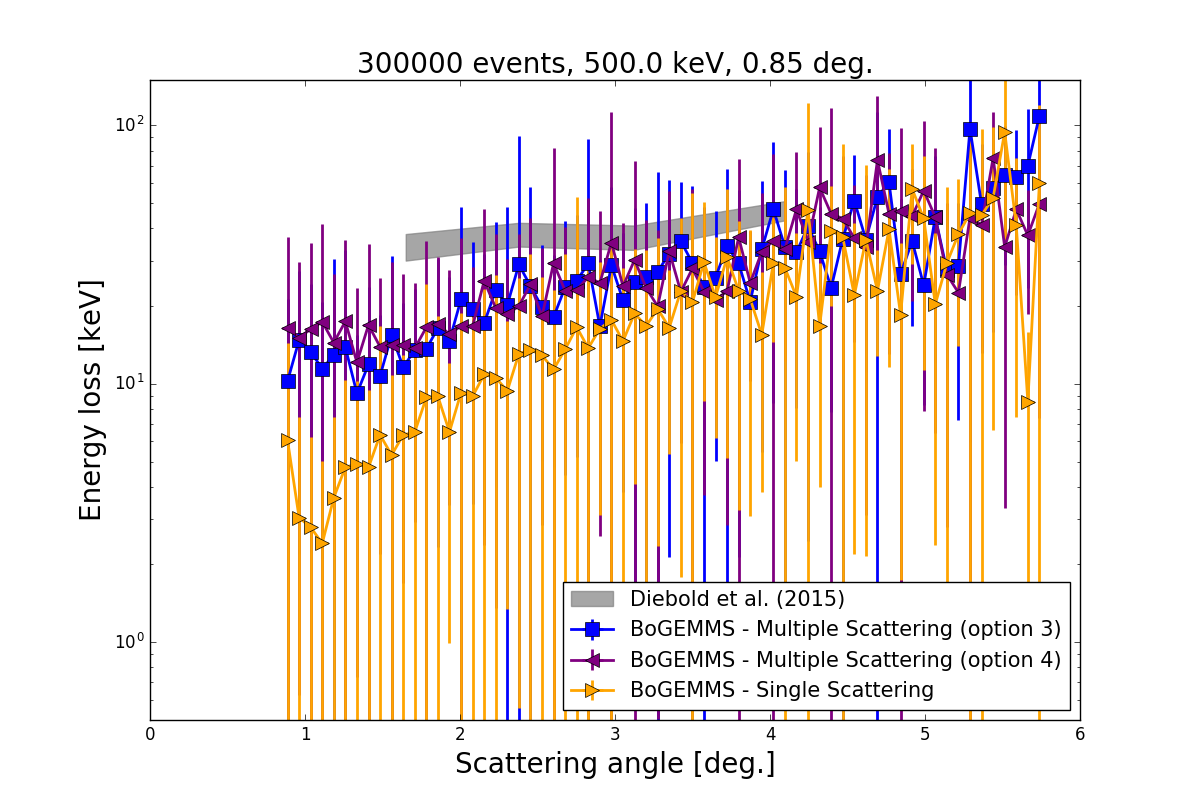}\\
    \includegraphics[width=0.49\textwidth]{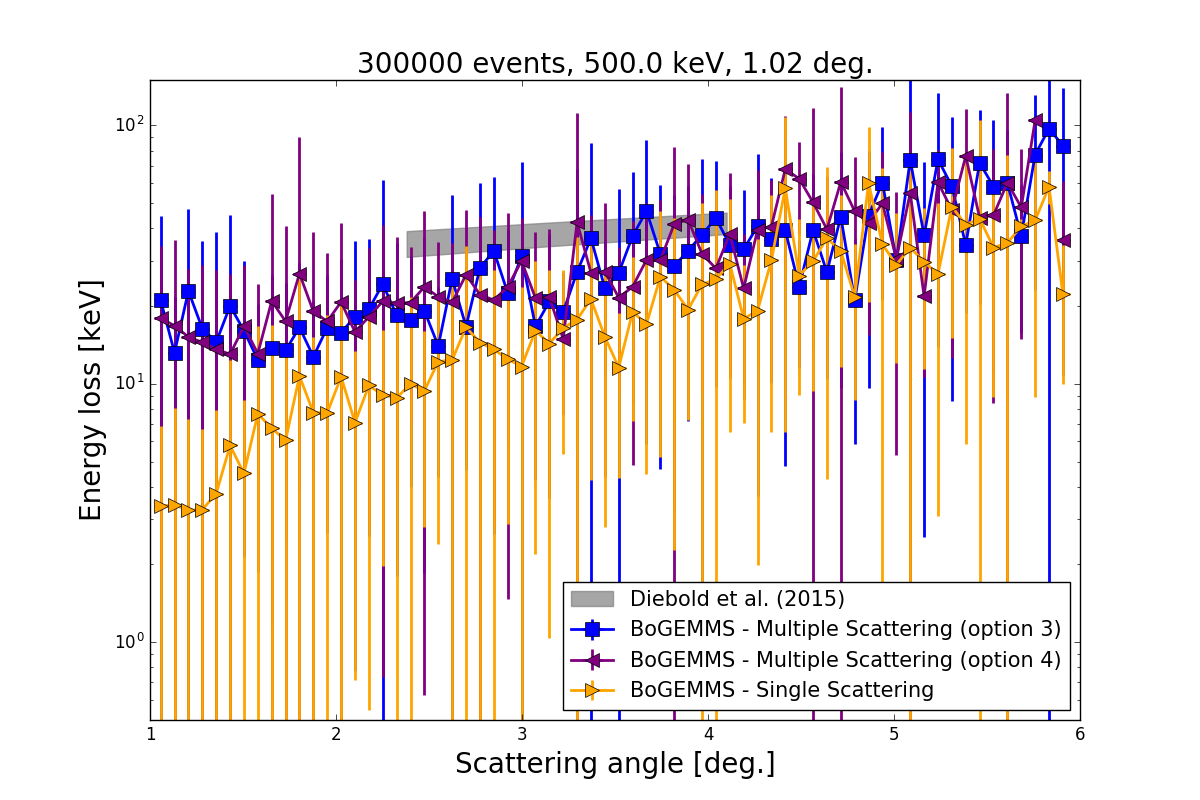}
  \includegraphics[width=0.49\textwidth]{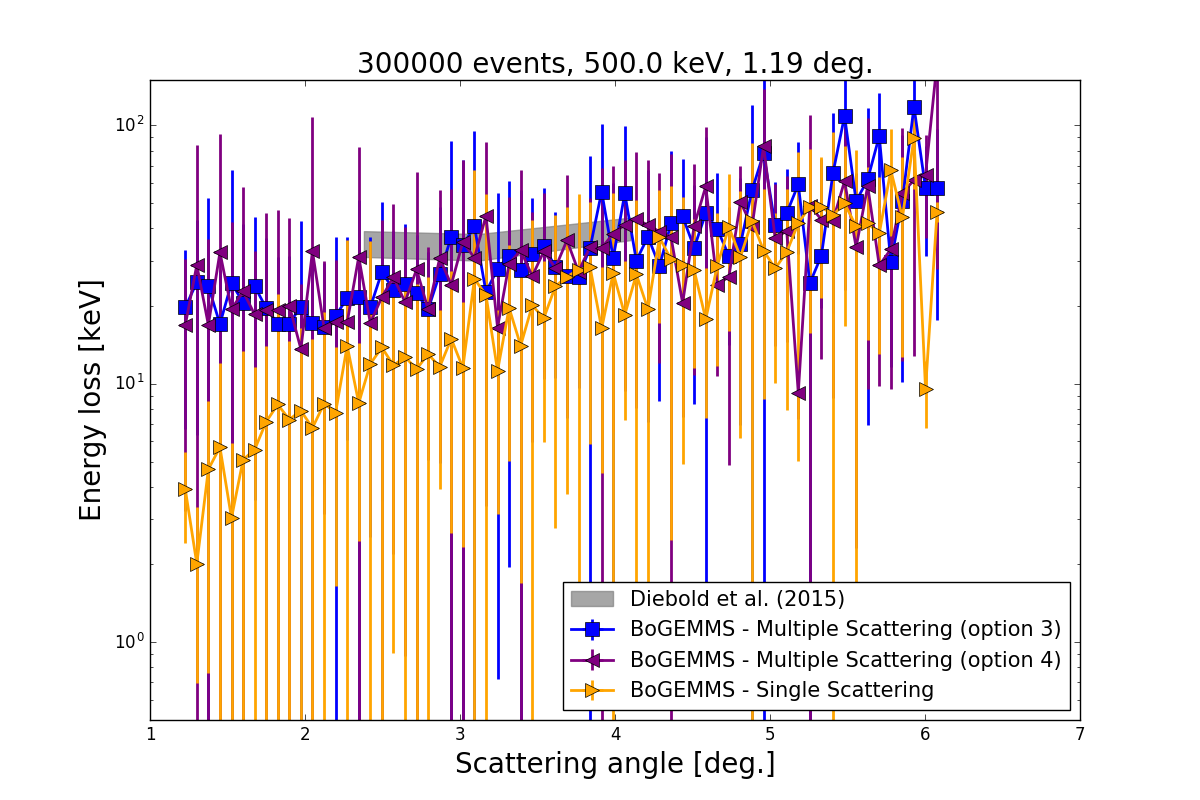}
\caption{The proton energy loss as a function of the scattering angle at E$_0$ = 500 keV in the $0.33^\circ - 1.19^\circ$ incident angle range. }
\label{fig:energy_500}     
\end{figure}
\begin{figure}[h!]
\center
  \includegraphics[width=0.49\textwidth]{energy_1000_1_label.png}
  \includegraphics[width=0.49\textwidth]{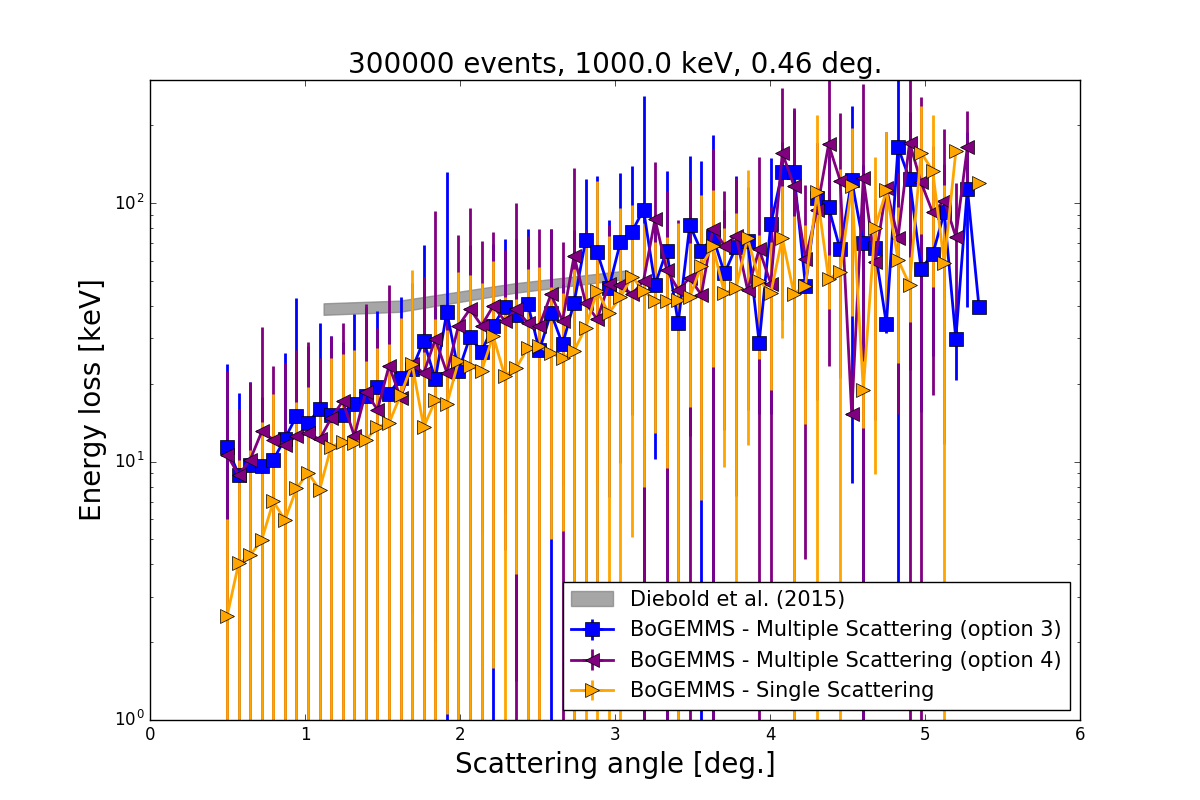}\\
  \includegraphics[width=0.49\textwidth]{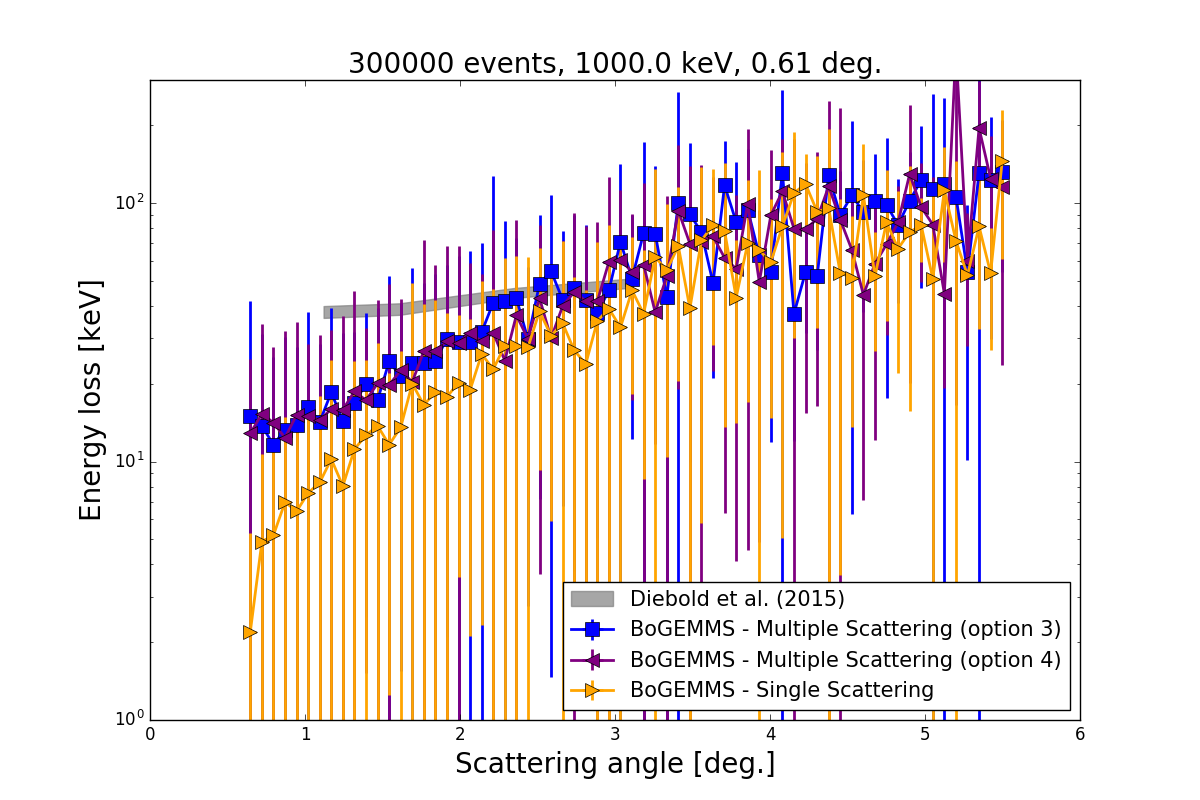}
  \includegraphics[width=0.49\textwidth]{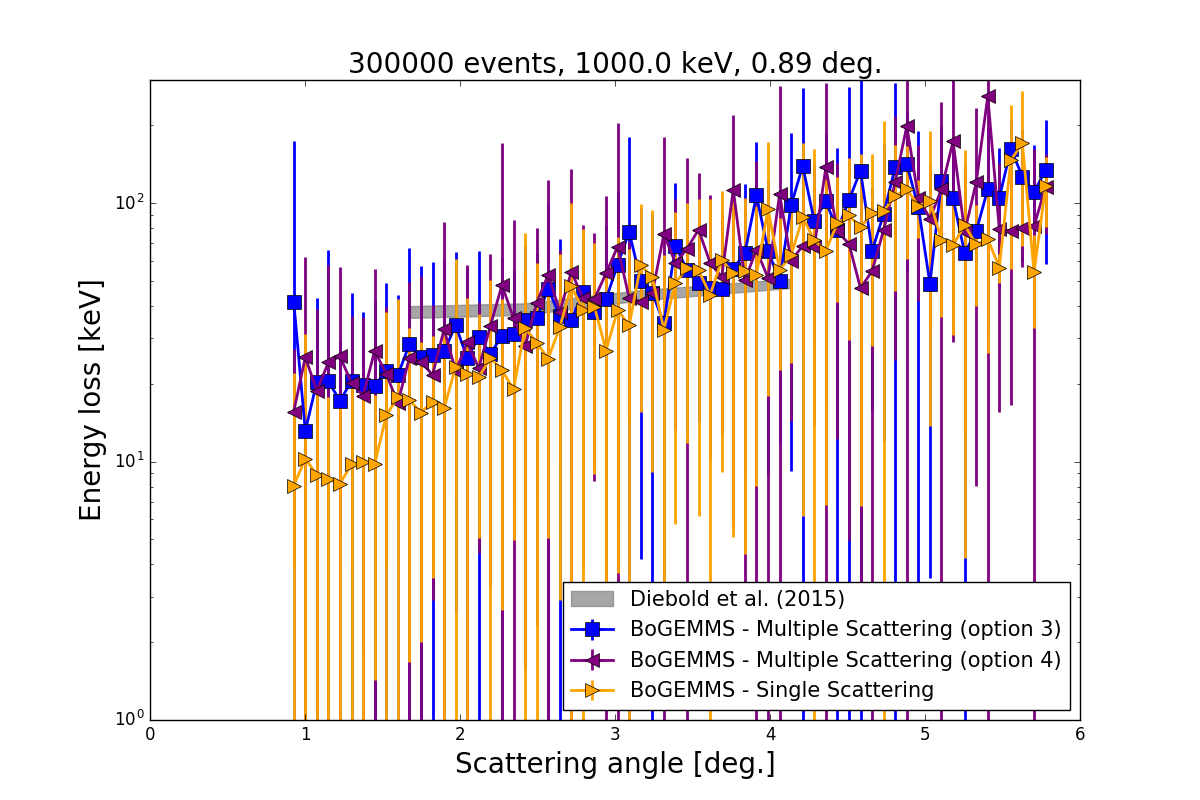}\\
    \includegraphics[width=0.49\textwidth]{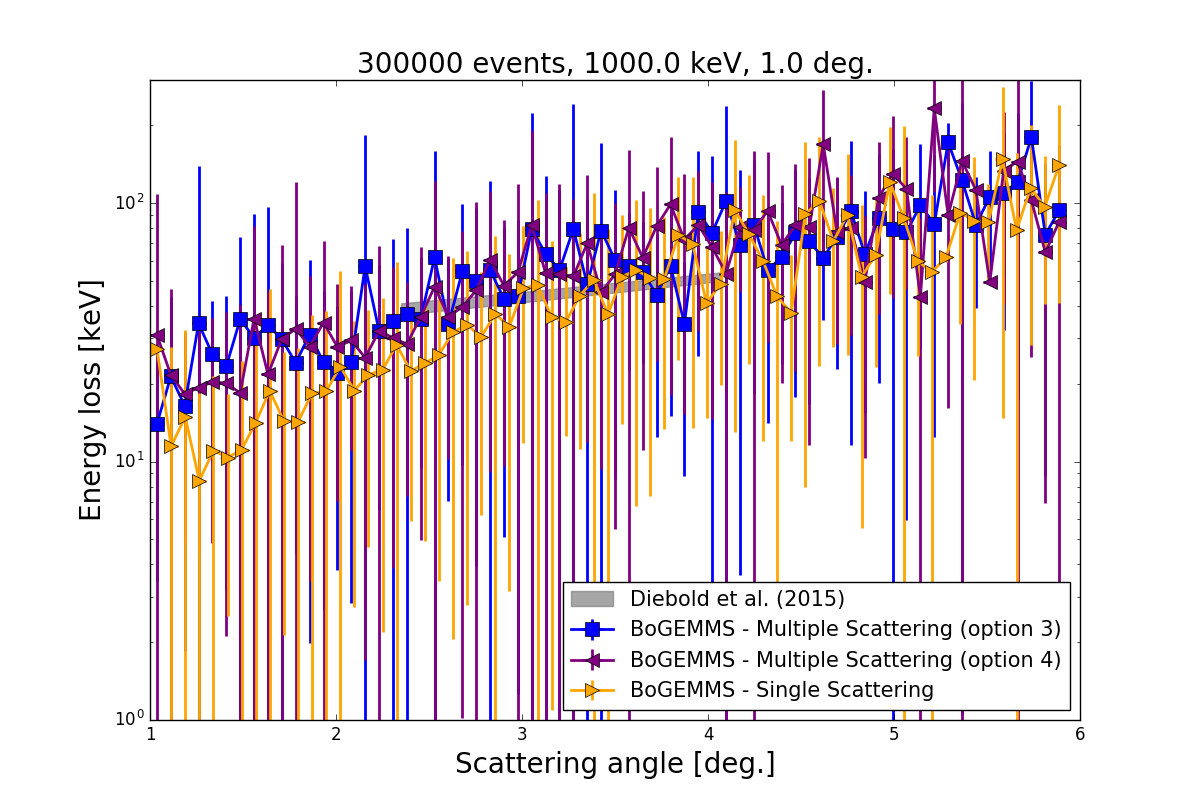}
  \includegraphics[width=0.49\textwidth]{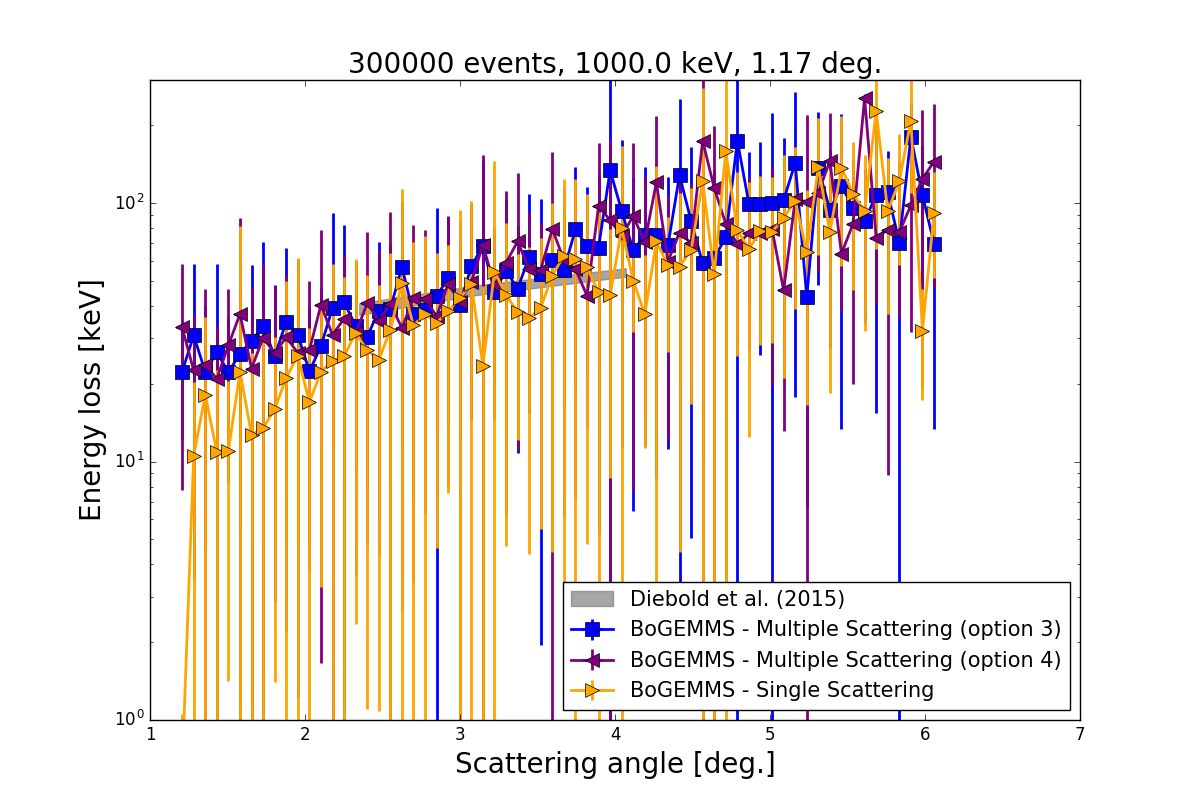}
\caption{The proton energy loss as a function of the scattering angle at E$_0$ = 1000 keV in the $0.3^{\circ} - 1.17^{\circ}$ incident angle range. }
\label{fig:energy_1000}     
\end{figure}

\end{document}